\documentclass[manuscript]{acmart}

\AtBeginDocument{%
  \providecommand\BibTeX{{%
    \normalfont B\kern-0.5em{\scshape i\kern-0.25em b}\kern-0.8em\TeX}}}

\setcopyright{acmcopyright}
\copyrightyear{2023}
\acmYear{2023}
\acmDOI{XXXXXXX.XXXXXXX}

\usepackage{todonotes}
\usepackage{csquotes}
\usepackage{siunitx}
\usepackage{soul} 
\usepackage{booktabs}

\usepackage{acro}
\acsetup{single}
\DeclareAcronym{aic}{short=AIC, long=Akaike information criterion}
\DeclareAcronym{bic}{short=BIC, long=Bayesian information criterion}
\DeclareAcronym{casa}{short=CASA, long=Computers As Social Actors}
\DeclareAcronym{cfa}{short=CFA, long=confirmatory factor analysis}
\DeclareAcronym{cfi}{short=CFI, long=comparative fit index}
\DeclareAcronym{efa}{short=EFA, long=exploratory factor analysis}
\DeclareAcronym{hci}{short=HCI, long=human-computer interaction}
\DeclareAcronym{hhd}{short=HHD, long=human-human dialogue}
\DeclareAcronym{hmd}{short=HMD, long=human-machine dialogue}
\DeclareAcronym{hri}{short=HRI, long=human-robot interaction}
\DeclareAcronym{ier}{short=IER, long=insufficient effort responding}
\DeclareAcronym{idaq}{short=IDAQ, long=Individual Differences in Anthropomorphism Questionnaire}
\DeclareAcronym{ipa}{short=IPA, long=Intelligent Personal Assistant}
\DeclareAcronym{kmo}{short=KMO, long=Kaiser-Meyer-Olkin}
\DeclareAcronym{mi}{short=MI, long=modification indices}
\DeclareAcronym{nasa-tlx}{short=NASA-TLX, long=NASA Task Load Index}
\DeclareAcronym{pca}{short=PCA, long=principal component analysis}
\DeclareAcronym{pmq}{short=PMQ, long=Partner Modelling Questionnaire}
\DeclareAcronym{rgt}{short=RGT, long=Repertory Grid technique}
\DeclareAcronym{rmsea}{short=RMSEA, long=root mean square error of approximation}
\DeclareAcronym{rmsr}{short=RMSR, long=root mean square of residuals}
\DeclareAcronym{sassi}{short=SASSI, long=Subjective Assessment of Speech System Interfaces}
\DeclareAcronym{sem}{short=SEM, long=structural equation modelling}
\DeclareAcronym{srmr}{short=SRMR, long=Standardised Root Mean Square Residual}
\DeclareAcronym{sus}{short=SUS, long=System Usability Scale}
\DeclareAcronym{tli}{short=TLI, long=Tucker-Lewis index}
\DeclareAcronym{icc}{short=ICC, long=Intra-class correlation coefficient}
\begin{document}

\title[Partner Modelling Questionnaire]{The Partner Modelling Questionnaire: A validated self-report measure of perceptions toward machines as dialogue partners}



\author{Philip R. Doyle}
\affiliation{%
  \institution{School of Information and Communication Studies, University College Dublin}
  \streetaddress{Belfield}
  \city{Dublin}
  \country{Ireland}}
\orcid{0000-0002-2686-8962}
\email{philip.doyle1@ucdconnect.ie}

\author{Iona Gessinger}
\affiliation{%
  \institution{School of Information and Communication Studies, University College Dublin}
  \streetaddress{Belfield}
  \city{Dublin}
  \country{Ireland}}
\orcid{0000-0001-5333-9794}
\email{iona.gessinger@ucd.ie}

\author{Justin Edwards}
\affiliation{%
  \institution{Learning and Educational Technology Research Lab, University of Oulu}
  \streetaddress{Belfield}
  \city{OUlu}
  \country{Finland}}
\orcid{0000-0003-1487-9207}
\email{jbardedwards@gmail.com}

\author{Leigh Clark}
\affiliation{%
  \institution{Computational Foundry, Swansea University}
  \streetaddress{}
  \city{Swansea}
  \country{Wales}}
\orcid{0000-0002-9237-1057}
\email{leighmhclark@gmail.com}

\author{Odile Dumbleton}
\affiliation{%
  \institution{School of Information and Communication Studies, University College Dublin}
  \streetaddress{Belfield}
  \city{Dublin}
  \country{Ireland}}
\orcid{0000-0003-3857-5099}
\email{odile.dumbleton@ucd.ie}

\author{Diego Garaialde}
\affiliation{%
  \institution{School of Computer Science, University College Dublin}
  \streetaddress{Belfield}
  \city{Dublin}
  \country{Ireland}}
\orcid{0000-0002-6034-2761}
\email{diego.garaialde@ucd.ie}

\author{Daniel Rough}
\affiliation{%
  \institution{School of Science and Engineering, University of Dundee}
  \streetaddress{Queen Mother Building}
  \city{Dundee}
  \country{Scotland}}
\orcid{0000-0003-1545-5377}
\email{drough001@dundee.ac.uk}

\author{Anna Bleakley}
\affiliation{%
  \institution{School of Information and Communication Studies, University College Dublin}
  \streetaddress{Belfield}
  \city{Dublin}
  \country{Ireland}}
\email{anna.bleakley@ucdconnect.ie}

\author{Holly P. Branigan}
\affiliation{%
  \institution{Psychology Department, University of Edinburgh}
  \streetaddress{7 George Square}
  \city{Edinburgh}
  \country{Scotland}}
\orcid{0000-0002-7845-8850}
\email{Holly.Branigan@ed.ac.uk}

\author{Benjamin R. Cowan}
\affiliation{%
  \institution{School of Information and Communication Studies, University College Dublin}
  \streetaddress{Belfield}
  \city{Dublin}
  \country{Ireland}}
\orcid{0000-0002-8595-8132}
\email{benjamin.cowan@ucd.ie}

\renewcommand{\shortauthors}{Doyle et al.}


\begin{abstract}
Recent work has looked to understand user perceptions of speech agent capabilities as dialogue partners (termed partner models), and how this affects user interaction. 
Yet, partner model effects are currently inferred from language production as no metrics are available to quantify these subjective perceptions more directly. 
Through three phases of work, we develop and validate the Partner Modelling Questionnaire (PMQ): an 18-item self-report semantic differential scale designed to reliably measure people's partner models of non-embodied speech interfaces. 
Through confirmatory factor analysis, we confirm that the PMQ scale consists of three factors: communicative competence and dependability, human-likeness in communication, and communicative flexibility. 
Our studies show that the measure consistently demonstrates good internal reliability, strong test-retest reliability over 4- and 12-week intervals, and predictable convergent/divergent validity. 
Based on our findings, we discuss the multidimensional nature of partner models, whilst identifying key future research avenues that the development of the PMQ facilitates. 
Notably, this includes the need to identify the activation, sensitivity, and dynamism of partner models in speech interface interaction. 
\end{abstract}



\maketitle

\section{Introduction}

Technologies that use speech as a primary interface modality are now commonplace. 
Yet we currently lack a detailed understanding of what drives people's dialogue behaviours in interactions with this type of interface \cite{clark2019state}.
Recent work aimed at cultivating this type of knowledge has focused on a concept known as `partner modelling' \cite{doyle2021we, brennan_two_2010}.
Partner models are said to reflect perceptions of a dialogue partner's communicative ability, and have been shown to influence language production in both \ac{hhd} and \ac{hmd} \cite{branigan_role_2011, cowan2015voice}, with people adapting their speech and language behaviours based on their partner model for both human and machine dialogue partners \cite{branigan_role_2011, cowan_what_2017}. 

Partner models in \ac{hmd} are assumed to depict machines as being comparatively basic and rigid, with limited conversational ability and competence \cite{branigan_role_2011}.
Elsewhere \cite{oviatt_linguistic_1998} speech interfaces are said to be seen by users as \emph{at-risk} listeners (i.e., dialogue partners with whom communication is at a high risk of failure). 
Concepts like audience design \cite {bell1984language} or recipient design \cite{an2021recipient}, similarly allude to partner models informing how speakers adapt the language they use when interacting with machines \cite{branigan_role_2011, brennan_two_2010}. 
This partner model driven adaptation in language  production is said to be triggered by a host of verbal and non-verbal cues such as an interlocutor's appearance, physical gestures, their accent, the language they use, \cite{branigan_role_2011, brennan_two_2010, nickerson1999we}, and/or events that occur during an interaction \cite{cowan_what_2017, brennan_two_2010}. 
In addition to influencing a person's language production in \ac{hmd} \cite{branigan_role_2011, cowan_whats_2019, an2021recipient}, partner models are also said to influence broader interaction behaviours by informing expectations about the types of tasks a machine dialogue partner might be capable of executing \cite{doyle_humanness_2019, cowan_what_2017, luger_like_2016}.
 
Currently, the role of partner models in language production is inferred rather than directly measured, with specific changes in language production (e.g., levels of lexical alignment -- see \cite{branigan_role_2011, cowan_whats_2019}) being interpreted as evidence of the use of partner models to guide interaction.  
However, findings using this approach have been somewhat inconsistent \cite{branigan_role_2011, pearson_adaptive_2006}, suggesting changes in language production might not always occur in conjunction with changes in partner models. 
Indeed, it has been suggested that, given the highly adaptive nature of language production, effects such as lexical alignment may be just one of a number of behaviours people adopt to achieve communicative success as a result of their beliefs about a partner's communicative abilities \cite{doyle2021we}.

Our work aims to build on recent development work \cite{doyle2021we} by validating a reliable subjective measure of people's partner models. More specifically, we report on the construction and validation of a self-report scale designed to measure the concept more directly, and independent of language behaviour.
The independence afforded by a subjective measure of partner models provides a significant contribution to the field, enabling researchers to quantify and attribute the consequences of design features and interaction events to changes in partner models, along with providing a way to independently assess whether changes in partner models for dialogue based systems map to changes in users' language production and broader interaction behaviours. 
Access to such a scale would also facilitate a more nuanced understanding of what constitutes a partner model in the context of \ac{hmd}. 
In other words, work here aims to give \ac{hmd} researchers a reliable, validated tool for detailed measurement of the potentially nuanced impact of specific design choices on user perceptions toward a speech interface system, along with a way to assess any potential impact this may have on interaction behaviour.  
Aside from being of benefit to researchers with an interest in partner modelling, a reliable validated scale of this nature, that is quick and relatively cheap to administer, would also benefit designers and user experience researchers, in commercial and academic settings. As such, the work addresses significant theoretical and methodological gaps both for researchers involved in work on partner modelling, and for researchers and developers of conversational user interfaces more broadly.

The development and validation of the \ac{pmq} is outlined through the findings of three phases of work, with each focusing on a key stage of scale construction and validation: initial item generation, scale construction, and identification of it's underlying factor structure using principal component analysis (\autoref{sec:phase1}, which is described in detail in \cite{doyle2021we}); evaluation of the scale's construct validity using confirmatory factor analysis (\autoref{phase2}); and evaluation of scale's convergent/divergent validity and test-retest reliability (\autoref{phase3}). 

In the first phase, focusing on scale construction, 390 participants responded to an initial battery of 51 items, with subsequent \ac{pca} leading to the identification of a 23-item questionnaire, with three underlying factors. These factors measured perceptions of \emph{communicative competence and dependability}, \emph{human-likeness in communication} and \emph{communicative flexibility}, while the scale and each of the factors also demonstrated high internal reliability. This phase is summarised so as to give readers a clear view of how items were generated and then selected in the initial construction and standardization of the scale. A detailed description of this phase can be found in \cite{doyle2021we}.

Phase 2 focused on confirmatory factor analysis of the scale's factor structure. Confirmatory factor analysis was performed on data from a study (study 1), where 117 participants were asked to complete the 23-item \ac{pmq} along with two further scales used to assess (i) speech interface usability (\ac{sassi}  \cite{hone2000towards}), and (ii) people's tendencies to engage in anthropomorphising behaviours towards non-human entities (\ac{idaq} \cite{waytz2010sees}). These additional subjective measures were included for assessment of convergent/divergent validity (\autoref{sec:convDiv}). 
Results from the \ac{cfa} suggested that the 23-item three factor structure was relatively robust, but that removal of five items to create an 18-item version may lead to improved stability. 

A further study (study 2) was then conducted, gathering a new sample of data (254 participants) with the aim of directly comparing the 23-item and 18-item models and identifying which had the better construct validity according to \ac{cfa} indices of good fit. This tests the hypothesis that the 18-item version of the scale offers the most reliable, robust and parsimonious account of partner models according to our data.
This study also adopted a repeated measures design, whereby participants from the same sample were asked to complete the \ac{pmq} again after a 12-week interval, and then again after a 4-week interval. This was to facilitate an evaluation of the scale's test-retest reliability (\autoref{sec:reTest}).
Results from the \ac{cfa} were consistent with those of the first study, suggesting that the 18-item three factor solution was the most parsimonious and robust version of the \ac{pmq}. 

Phase three of the work focused on assessing the validity and test-retest reliability of the measure developed. Analysis of study 1 data showed that the 18-item \ac{pmq} demonstrated good convergent/divergent validity, correlating as predicted with certain \ac{sassi} subscales and not correlating, as predicted, with other \ac{sassi} subscales or the \ac{idaq} scale. Results suggest that the \ac{pmq} performs as predicted compared to the \ac{sassi} and \ac{idaq} scales, and that it measures something distinct compared to these established psychometric tools. 

Test-retest reliability for the 18-item measure, assessed using \ac{icc} and Pearson's correlations, also proved to be relatively strong at both 12- and 4-week intervals; particularly considering the nature of the scale and the broad psychological nature of the construct it was designed to address \cite{watson2013effect, kline_handbook_2013}. The final 18-item scale is included in \autoref{tab:final18} with scoring and administration instructions included in supplementary material. All data and analysis scripts for studies 1, 2a and b within this work are available at \url{https://osf.io/xnvry/?view_only=d3ad5f6dd24549469eabd5ae0dfe5ab5}. 

\begin{table}
\centering
\caption{Final 18 item version of the Partner Modelling Questionnaire. Factor 1: Competence and Dependability; Factor 2: Human-Likeness; Factor 3: Communicative Flexibility}
\label{tab:final18}
\begin{tabular}{ll}
\toprule
Item & {Subscale}\\
\midrule
Competent -- Incompetent & \\
Dependable -- Unreliable & \\
Capable -- Incapable & \\
Consistent -- Inconsistent & Competence and Dependability\\
Reliable -- Uncertain & \\
Expert -- Amateur & \\
Efficient -- Inefficient & \\
Precise -- Vague & \\
Cooperative -- Uncooperative & \\
\midrule
Human-like -- Machine-like  & \\
Life-like -- Tool-like  & \\
Warm -- Cold  & Human-Likeness\\
Empathetic -- Apathetic  & \\
Personal -- Generic  & \\
Social -- Transactional  & \\
\midrule
Flexible -- Inflexible  & \\
Interpretive -- Literal & Communicative Flexibility\\
Spontaneous -- Predetermined  & \\
\bottomrule
\end{tabular}
\end{table}

\section{Related work}


\subsection{The concept of partner models}
\label{pm_concept}

The basic tenet of partner modelling is that people form a mental representation of their dialogue partner as a communicative and social entity \cite{branigan_role_2011,cowan_what_2017}. 
Originating in psycholinguistics, the concept proposes that this mental representation informs what people say to a given interlocutor, how they say it, and the types of tasks someone might entrust their partner to carry out \cite{brennan_two_2010, branigan_role_2011}. 
Hence, partner models might also be understood as a heuristic account of a partner's communicative ability and social relevance that guides a speaker toward interaction and language behaviours that are appropriate for a given interlocutor. 
In this sense, it is similar to accounts of mental models in cognitive psychology \cite[e.g.,][]{johnson-laird_mental_1980, johnson-laird_mental_2010} and Norman's explanation of mental models in \ac{hci} \cite{norman1983some}.
Indeed, a partner model can be broadly understood as a mental model of the dialogue partner. Informed by research examining the role of partner models in \ac{hhd} and \ac{hmd} interactions \cite{branigan_role_2011, brennan_two_2010, clark_using_1996, cowan2015voice}, and established explanations of closely related concepts such as mental models \cite{johnson-laird_mental_1980, johnson-laird_mental_2010, norman1983some, westbrook_mental_2006} and theory of mind \cite{baron-cohen_friendship_2003, baron1999evolution, byom2013theory}, recent work proposed a working definition of partner models, defining them as: \emph{"an interlocutor’s cognitive representation of beliefs about their dialogue partner’s communicative ability. These perceptions are multidimensional and include judgements about cognitive, empathetic and/or functional capabilities of a dialogue partner. Initially informed by previous experience, assumptions and stereotypes, partner models are dynamically updated based on a dialogue partner’s behaviour and/or events during dialogue"} \cite{doyle2021we}


In psycholinguistics, it is assumed that partner models are initially derived from minimal cues such as accent, age, or appearance \cite{clark_using_1996, nickerson1999we, branigan_role_2011}. 
These models are thought to inform people's language production in dialogue, leading them to design utterances around their audience's perceived capabilities and social background with the aim of increasing the chance of communicative success. In this sense they incorporate the notion of audience design \cite{bell1984language, brennan_conceptual_1996, branigan_role_2011}. 
Indeed, audience design effects are seen in \ac{hmd}, whereby cues such as accent and a system's projected nationality have been shown to prompt users to adopt congruent lexis \cite{cowan_whats_2019}. 
Work has also proposed that, when compared to human dialogue, people see computers as basic or at-risk dialogue partners, leading them to simplify syntactic and lexical choices \cite{branigan_role_2011, oviatt_linguistic_1998}. 
Collectively, the work suggests that both social and competence related factors contribute to partner models in the context of \ac{hmd}.

\ac{hhd} work on partner modelling also suggests that, rather than being static, partner models may be dynamic in nature. 
Specifically, it is proposed that initially people use \emph{global partner models}, which are derived both from past experiences with seemingly similar dialogue partners and subtle cues perceived early in an interaction.
However, when beneficial to do so, people work toward developing a more accurate, individualised \emph{local partner model} for a specific dialogue partner based on their interaction experiences with that partner \cite{brennan_two_2010}.



\subsection{Measuring partner models in \ac{hmd}}
\label{sec:partnerModelMeasurement}
Rather than measuring the concept of partner models directly, current research in this area assumes a partner modelling effect through changes in linguistic choices when partner characteristics, capabilities or dialogue events are experimentally manipulated. 
In particular, alignment -- the tendency for interlocutors to converge on lexical, syntactic, and/or prosodic features of speech during dialogue -- is thought to be sensitive to partner modelling \cite{branigan_linguistic_2010, brennan_two_2010, cowan_what_2017, pearson_adaptive_2006}.
Specifically, stronger levels of alignment (a.k.a.\ entrainment, accommodation, convergence) have been shown in situations where communication is perceived to be at an increased risk of failure due to an interlocutor’s perceived communicative limitations \cite{branigan_linguistic_2010, pearson_adaptive_2006}. 
In \ac{hhd}, this effect has been shown in dialogue between experts and novices \cite{isaacs_references_1987}, first- and second-language speakers \cite{bortfeld1997use}, and when adults speak to children or infants \cite{misiek_development_2020}. 
In \ac{hmd}, work has largely focused on lexical alignment -- convergence on word choices -- which seems particularly sensitive to a dialogue partner’s perceived communicative limitations, with significantly higher levels of lexical alignment seen in \ac{hmd} compared to \ac{hhd} \cite{bergmann_exploring_2015, branigan_role_2011}.
These effects are interpreted as being indicative of a user holding a poor partner model of a machine as a dialogue partner. 
Recent research on lexical choices also shows similar effects, akin to audience design \cite{bell1984language}, whereby participants who interacted with a machine dialogue partner that had been designed to project an American persona tended to use more US lexical terms compared to interaction with a system designed to project an Irish persona. 
What is more, research on \ac{hmd} consistently highlights how linguistic adaptation within interaction may be driven by people's partner models, with people changing grammatical structure or hyper-articulating \cite{porcheron2018voice} based on a perception of a machine dialogue partner as an \emph {at-risk listener} \cite{oviatt1998predicting}. 
Notably, these effects are not consistently identified, with work using a similar experimental paradigm finding no effect of partner models on alignment \cite{cowan_does_2015,cowan2015voice}. In sum, current efforts to assess partner models tend to focus on their effects, particularly on language production, rather than measuring the concept explicitly.

\subsection{Subjective measures in \ac{hmd} research}

Subjective measures are used frequently in \ac{hmd} research, normally to investigate the impact of system design features on usability and user attitudes such as perceptions of trust and likeability, user preference and technology acceptance \cite{clark2019state, seaborn_measuring_2021}. 
However, although subjective measures are widely deployed, questions have been raised about the reliability and validity of the largely bespoke questionnaires normally used in these studies \cite{clark2019state, seaborn_measuring_2021, bruggemeier_user_2020, kocaballi_understanding_2019}. 
For instance, a review of 68 speech interface studies \cite{clark2019state} found that out of the 40 papers that used questionnaires to assess user attitudes, 38 were custom built, with many using single items to measure concepts, which has been noted as unreliable \cite{cairns_commentary_2013, kline_handbook_2013}. 
In terms of standardised scales, the review by Clark et al.~\cite{clark2019state} identified: only one use of the \ac{sassi} questionnaire \cite{hone2000towards} designed to measure usability in the context of \ac{hmd}; three uses of the \ac{sus} \cite{brooke_sus_1996} designed to measure \ac{hci} user experience more broadly; and three uses of the \ac{nasa-tlx} \cite{hart_development_1988} designed to measure mental workload, a concept closely related to cognitive load. 
Similar observations have been made across several recent reviews of \ac{hmd} research \cite{bruggemeier_user_2020, kocaballi_understanding_2019, seaborn_voice_2021, seaborn_measuring_2021}. 
The lack of reliable, validated subjective measures developed within the context of \ac{hmd} \cite{bruggemeier_user_2020, clark2019state, seaborn_measuring_2021} almost certainly plays a role in why researchers commonly choose to use bespoke scales. 
However, in addition to concerns about the statistical reliability of findings, failure to develop reliable, validated questionnaires also leads to consequences in terms of generalisability, as even if a selection of studies are focused on the same concept, there is little continuity in terms of how the concept is understood and quantified \cite{kocaballi_understanding_2019, seaborn_voice_2021}.

\subsection{Research aims}

As previously highlighted, partner models are assumed to impact user behaviour in human machine dialogue. 
Yet currently we have no way of assessing partner models, other than inferring their impact through behavioural studies of language production. 
This leaves us both unable to measure a person's partner model for a given partner, nor assess how this model may change based on dialogue experiences or changes to system design features. 
Our work looks to contribute to this body of research by developing and validating a subjective measure of partner models in \ac{hmd} interaction that provides a greater level of detail in terms of understanding the factors that contribute to partner models for machine dialogue partners through a clearly defined, reliable and valid psychometric tool. 
The scale may prove useful for understanding the impact of system design on user perceptions, and allow for examination of how changes to one design feature might have unintended consequences elsewhere. 
This paper describes three separate phases of work conducted to first develop and then validate the scale.
The first phase, described in detail in earlier work \cite{doyle_humanness_2019, doyle2021we}, outlines initial exploratory work carried out during the construction of a 23-item questionnaire using \ac{pca}. 
Some explanation on this is included here for completeness.
Within the second phase of work, we ran a study (study 1), a replication of the first, to test the questionnaire's construct validity (i.e., to confirm the scale's hypothesised structure) through confirmatory factor analysis (CFA). This analysis failed to uphold the hypothesised structure, but guided us toward a reduced 18-item model that may meet \ac{cfa} indices of good fit.
A second study (study 2) was then run with the aim of directly comparing the 23- and 18-item versions, and ultimately attest to the 18-item version's construct validity.
Phase 3 revolves around further validation and reliability tests. During study 1, we also administered two associated scales to assess the \acp{pmq} convergent and divergent validity, whilst in study 2 we administered the PMQ across two time intervals with the aim of evaluating the \acp{pmq} test-retest reliability at 4- and 12-week intervals.

\section{Phase 1: Item generation and scale construction}
\subsection{Item pool generation and reduction}
\label{sec:phase1}
Before outlining the studies that were conducted to validate the \ac{pmq}, we first provide some background detail on the origin of scale items and the initial factor structure of the scale. 
Item generation involved a complex multistage process involving three distinct stages. 
These stages are described in greater detail in \cite{doyle2021we, doyle_humanness_2019}, however, for the sake of completeness the process and outcome of each stage is briefly described below.

\subsubsection{Stage 1 -- Repertory Grid Study}

The first step in this process focused on generating potential items from participants based on their experiences interacting with speech interfaces. 
These were generated as part of a study using the \ac{rgt} \cite{kelly1970brief} wherein 21 participants were asked to generate word pairs (termed `personal constructs') that describe key similarities and differences between an object of interest and appropriate comparators (termed `elements'). 
In this case, the comparators, or elements participants were asked to interact with included three different types of dialogue partner: two speech interfaces (Apple's Siri, which was accessed through a smartphone, and Amazon Alexa, which was accessed through an Echo Dot smart speaker), and a human (a member of the research team). Participants asked each of the partners as set of nine questions that were either conversational (e.g., Where are you from?), information retrieval (e.g., How do I get to the city centre from here?) or subjective/opinion based (e.g., What do you think of [insert famous person's name]?) questions.
After this, participants were then asked to generate a list of words (termed `implicit constructs') that they felt appropriately describe key similarities and differences between the elements \cite{jankowicz_easy_2004, fransella2004manual}. 
Using a think-aloud protocol, participants were also asked to explain what it is about the interaction that the implicit construct defines. 
Finally, participants were asked to generate a word to act as an opposite pole (termed `emergent construct') for each of the implicit constructs identified \cite{jankowicz_easy_2004, fransella2004manual}. 
This process generated 246 unique word pairs out of 266 in total. 
As mentioned, further details on the process and all words generated are included in \cite{doyle2021we}. 
For completeness, the full list of word pairs is included in the supplementary material of \cite{doyle2021we}.

\subsubsection{Stage 2 -- Systematic review of subjective measures}

The second step in item generation involved a systematic review of relevant subjective measures with the aim of identifying potential additional items to ensure the item pool provided adequate coverage \cite{kline_handbook_2013}. 
In total, a list of 75 potentially relevant measures was generated, with 44 identified as containing items to be included in the initial item pool after close review. 
A full list of questionnaires and the co-opted items are provided in the supplementary material of \cite{doyle2021we}. A more detailed account of this stage is presented in \cite{doyle2021we}.

\subsubsection{Stage 3 -- Item pool screening}

Two screening phases were then conducted with the aim of reducing the large pool of items generated. 
The challenge here was to reduce the item pool to include only items that were relevant to the concept of partner modelling (i.e., eliminate items with poor face validity), whilst retaining as much of the nuance and depth contained within the item pool as possible \cite{field_discovering_2013, kline_psychometrics_2000, kline_handbook_2013}. The initial screening phase involved screening of items generated from the  \ac{rgt} study and the subjective measure review. We focused on \ac{rgt} items first so as to prioritise user-generated items within scale development. \ac{rgt} item screening by the lead author resulted in a reduction from 266 to 88 items.  
Based on the form of these items (i.e. items including two opposite poles), and after assessing the benefits (i.e. rapid intuitive responding) and drawbacks (i.e. potential reduction in reliability due to increased openness to subjective interpretation)\cite{kline_handbook_2013} it was decided to adopt a semantic differential scale structure for the PMQ. Advice suggests that either a 7- or 9-point scale were most reliable whilst being more suitable for statistical analysis than shorter response scales (e.g., a 5-point scale) \cite{kline_handbook_2013}. 
Hence, it was decided to use the more common 7-point response scale.
Screening of items from existing subjective measures, resulted in 39 items being identified as candidates to be retained. All 39 items took the form of a short phrase, and therefore required adjective extraction and generation of an appropriate antonym. 
The result of this initial screening was a reduction in the total item pool from 266 to 127 items. Given this is still too many items to present to participants in the form of a questionnaire, a further screening phase was carried out, which involved an expert review that was also aimed at further reducing the item pool whilst retaining good face validity and coverage. This led to identification of a set of 51 items which were included in the initial draft of the \ac{pmq}. Further details about the review phase are process outlined in \cite{doyle2021we}.

\label{sec:study1}
\subsection{Initial factor structure}
The next step involved presenting the initial 51-item \ac{pmq} to participants with the aim of identifying its underlying factor structure and identify the item set that best represents that structure. This work aimed to take a quantitative data-driven approach to scale construction, with decisions about which items to retain/eliminate being based on assessment of inter-item correlations, factor loadings and communalities. Other work has also used steps like \emph{cognitive interviewing} \cite{willis2005cognitive, buers/etal:2014, jacobs/etal:2023} to support the removal of items earlier on in the item generation process, which could also be considered by those developing further measures within the field.
Again, a detailed description of this process can be found in \cite{doyle2021we}. 
The study summary is reported here for completeness. 
\begin{table}
\centering
\caption{3-factor model with factor loadings, shared and unique variation (h2, u2) and communality (com) after pruning (23 items)- taken from \cite{doyle2021we}}
\label{tab:pca23}
\begin{tabular}{l*{6}{S[table-format=1.2]}}
\toprule
Item & {Factor 1} & {Factor 2} & {Factor 3} & {h2} & {u2} & {com} \\
\midrule
Competent/Incompetent & 0.77 & & & 0.64 & 0.36 & 1.1 \\
Dependable/Unreliable & 0.68 & & & 0.54 & 0.46 & 1.2 \\
Capable/Incapable & 0.68 & & & 0.58 & 0.42 & 1.2 \\
Consistent/Inconsistent & 0.67 & & & 0.46 & 0.54 & 1.1 \\
Reliable/Uncertain & 0.67 & & & 0.51 & 0.49 & 1.2 \\
Clear/Ambiguous & -0.65 & & & 0.44 & 0.56 & 1.3 \\
Direct/Meandering & 0.64 & & & 0.43 & 0.57 & 1.4 \\
Expert/Amateur & 0.64 & & & 0.46 & 0.54 & 1.1 \\
Efficient/Inefficient & 0.64 & & 0.31 & 0.53 & 0.47 & 1.5 \\
Honest/Misleading & 0.63 & & -0.35 & 0.50 & 0.50 & 1.7 \\
Precise/Vague & 0.62 & & & 0.41 & 0.59 & 1.2 \\
Cooperative/Uncooperative & 0.54 & & & 0.44 & 0.56 & 1.5 \\
Human-like/Machine-like & & 0.75 & & 0.54 & 0.46 & 1.0 \\
Life-like/Tool-like & & 0.75 & & 0.58 & 0.42 & 1.0 \\
Warm/Cold & & 0.67 & & 0.48 & 0.52 & 1.2 \\
Empathetic/Apathetic & & 0.65 & & 0.43 & 0.57 & 1.1 \\
Personal/Generic & & 0.62 & & 0.43 & 0.57 & 1.0 \\
Authentic/Fake & & 0.56 & & 0.40 & 0.60 & 1.2 \\
Social/Transactional & & 0.54 & & 0.45 & 0.55 & 1.5 \\
Flexible/Inflexible & & & 0.66 & 0.54 & 0.46 & 1.2 \\
Interactive/Stop-Start & & & 0.61 & 0.50 & 0.50 & 1.4 \\
Interpretive/Literal & & & 0.56 & 0.47 & 0.53 & 1.5 \\
Spontaneous/Predetermined & & & 0.51 & 0.43 & 0.57 & 1.8 \\
\midrule
\multicolumn{1}{r}{SS loadings} & 5.46 & 3.56 & 2.18 & & & \\
\multicolumn{1}{r}{Proportion Var} & 0.24 & 0.15 & 0.09 & & & \\
\multicolumn{1}{r}{Cumulative Var} & 0.24 & 0.39 & 0.49 & & & \\
\multicolumn{1}{r}{Proportion Explained} & 0.49 & 0.32 & 0.19 & & & \\
\midrule
\multicolumn{1}{r}{Component correlations} & & & & & & \\
\multicolumn{1}{r}{Factor 1} & 1 & 0.22 & 0.11 & & & \\
\multicolumn{1}{r}{Factor 2} & 0.22 & 1 & 0.36 & & & \\
\multicolumn{1}{r}{Factor 3} & 0.11 & 0.36 & 1 & & & \\
\midrule
\multicolumn{7}{l}{Mean item complexity $= 1.3$} \\
\multicolumn{7}{l}{Test of the hypothesis that 3 components are sufficient.} \\
\multicolumn{7}{l}{RMSR $= 0.05$ (empirical $\chi^2 = 542.03$ with prob $2.8 \times 10^{-36}$)} \\
\multicolumn{7}{l}{Fit based upon off diagonal values $= 0.96$} \\
\bottomrule
\end{tabular}
\end{table}


390 participants (of which data from 356 participants were included in the final analysis after data removal) were asked to complete the 51-item PMQ  by
reflecting on the speech interfaces they interacted with most frequently and then respond to each item based on these
experiences. Specifically, participants were asked: “Thinking about the way speech interfaces communicate with you,
how would you rate their ability on a scale between each of the following poles?”. They then indicated where they
felt their preferred speech interface sat between each pole using the 7-point semantic differential scale. Participants
were also instructed to read each pair of words carefully, to respond as quickly and accurately as possible, and to
avoid providing too many neutral responses. Principal component analysis (PCA) was then conducted on the sample's responses to identify
the underlying factor structure and items that best represent those factors. In other words, with the aim of constructing a standardised questionnaire. First items were eliminated based on poor inter-item correlations (to low or too high). Next a 3-factor solution was identified based on eigenvalues and parallel analysis. Then further item reduction was carried out by iteratively pruning of items with low factor loadings and communalities, ultimately leading to a three-factor, 23-item model. This model accounted for 49\% of the variance in the original data, similar to the acceptable threshold for proportion of variance explained in samples of 100-200 participants \cite{MacCallum1999FAsampleSize, clark1995constructing}. Results of the PCA with factor loadings and communalities are included in \autoref{tab:pca23}. 

Naming conventions for questionnaire construction determine that factors are named based on items contained within a factor, with emphasis given to items with the highest factor loadings. With this in mind, authors discussed the most appropriate name for each factor based on the items with the highest factor loadings within them. We then discussed whether this name adequately represented the remaining items within each respective factor also. 
As such, \ac{pmq} factors appear to represent perceptions of \emph{communicative competence and dependability} (Factor 1: 12-items; $\alpha$=0.88, accounting for \SI{49}{\percent} of the variance within the model), perceptions of \emph{human-likeness in communication} (Factor 2: 7-items; $\alpha$=0.8, accounting for \SI{32}{\percent} of the variance within the model), and perceptions of \emph{communicative flexibility} (Factor 3: 4-items; $\alpha$=0.72, accounted for \SI{19}{\percent} of the variance within the model) respectively. A more detailed description of the procedure, data cleaning and analysis process can be found in \cite{doyle2021we}.

\section{Phase 2: Construct Validity}
\label{sec:study2}
\label{phase2}
We follow up work in phase one, described above, with two further studies (Study 1 and Study 2) aimed at assessing the construct validity of the proposed three-factor structure using confirmatory factor analysis (\ac{cfa}). In this section we report the results of a \ac{cfa} of the 23-item, three-factor structure (Study 1) and a follow-up \ac{cfa} to aimed at comparing  the original 23-item version with a more robust 18-item version, which was surfaced during Study 1 (Study 2). 
When conducting \ac{cfa}, a new sample of data is required, as confirming a factor structure based on the same data used to identify a model would create a confirmation bias \cite{brown_confirmatory_2015}.
For this initial \ac{cfa} (study 1), we recruited 135 participants who had volunteered to take part in previous work \cite{doyle2021we}. 
The study was conducted over a period of 12-18 weeks after this initial research. 

Study 1 involved participants completing the 23-item \ac{pmq} alongside two further questionnaires that were designed to measure related concepts. Namely, perceptions of speech interface usability (measured using the \ac{sassi} \cite{hone2000towards} scale), and general tendencies to anthropomorphise non-human objects (measured using the \ac{idaq} scale \cite{waytz2010sees}).
This facilitated evaluation of the \ac{pmq}'s convergent/divergent validity (i.e., its relationship to scales designed to measure similar concepts). 
Results from analysis of relationships between the PMQ, SASSI and IDAQ questionnaires are reported in \autoref{sec:convDiv}. 
The following sections focus instead on the results of the \ac{cfa} to evaluate the statistical construct (or structural) validity of the 23 and 18-item versions of the \ac{pmq} scales.

\subsection{Study 1 Participants}
\label{sec:participants2}

In total, 135 participants were recruited from participants who took part in a previous study \cite{doyle2021we}, 12 to 18 weeks earlier. 
Potential to participate in this follow up study was signalled to participants during debriefing for the initial study, with all participants being told they would be awarded two further entries into a €200 voucher prize draw (used as an incentive in the earlier work), if they also completed the current study. 
The study was given low-risk ethical clearance by University College Dublin's Human Research Ethics Committee (H-REC), and was conducted according to guidelines form the university's and the British Psychological Society. 

Data from 18 participants were excluded from analysis due to heavily patterned responses that lacked variation {(see \ref{sec:approach_cfa_study1}), which is seen as evidence of \ac{ier} \cite{maniaci2014caring, steedle2018detecting}} or due to failure to complete the survey which, in accordance with ethics approval, was treated as a withdrawal with data being deleted. 
The total sample therefore included 117 native or near-native English-speaking participants (63 female, 52 male, 2 preferred not to say; age range: 18-57 yrs, mean age=27yrs, SD age= 9.4 yrs). 

Details of the demographics of the sample for this study are included in \autoref{tab:education_combined} (educational levels), \autoref{tab:speech_interfaces_combined} (most frequently used speech interfaces), \autoref{tab:usage_frequency_combined} (usage frequency of speech interfaces), and \autoref{tab:device_types_combined} (devices used to access speech interfaces).

\begin{table}
    \centering
    \caption{Education levels of participants (\%) across studies.}
    \label{tab:education_combined}
    \begin{tabular}{lSSSS}
        \toprule
        \textbf{Education Level} & \textbf{Study 1} & \textbf{Study 2a} & \textbf{Study 2b} \\
        \midrule
        Secondary/Vocational & 35.0 & 43.3 & 43.4 \\
        Undergraduate & 30.8 & 37.8 & 40.6 \\
        Graduate/Postgraduate & 34.2 & 18.1 & 14.2 \\
        Other/Prefer not to say & {--} & 0.8 & 1.9 \\
        \bottomrule
    \end{tabular}
\end{table}

\begin{table}
    \centering
    \caption{Most frequently used speech interfaces (single response option choice) among participants (\%).}
    \label{tab:speech_interfaces_combined}
    \begin{tabular}{lSSSS}
        \toprule
        \textbf{Speech Interface} & \textbf{Study 1 } & \textbf{Study 2a} & \textbf{Study 2b} \\
        \midrule
        Apple Siri & 36.8 & 27.6 & 23.6 \\
        Google Assistant & 33.3 & 25.6 & 29.2 \\
        Amazon Alexa & 25.6 & 43.3 & 44.3 \\
        Microsoft Cortana & 2.6 & 1.2 & 1.0 \\
        Other & 1.7 & 2.0 & 1.0 \\
        \bottomrule
    \end{tabular}
\end{table}

\begin{table}
    \centering
    \caption{Frequency of speech interface usage among participants across studies. For study 1, ratings range from 1 (very infrequently) and 7 (very frequently); for studies 2a and 2b, results are given in \% per category.}
    \label{tab:usage_frequency_combined}
    \begin{tabular}{lSS}
        \toprule
        \textbf{Usage Frequency (1-7)} & \ \textbf{Study 1} \\
        \midrule
        Mean & 4.0 \\
        SD & 1.9 \\
        \midrule

        \textbf{Usage Frequency} & \textbf{Study 2a} & \textbf{Study 2b} \\
        \midrule    
        Rarely & 14.2 & 12.3 \\
        Few times a month & 16.5 & 17.9 \\
        Few times a week & 29.9 & 31.1 \\
        Daily & 39.4 & 38.7 \\
        \bottomrule
    \end{tabular}
\end{table}

\begin{table}
    \centering
    \caption{Devices used to access speech interfaces among participants (\%) across studies.}
    \label{tab:device_types_combined}
    \begin{tabular}{lSSSS}
        \toprule
        \textbf{Device Type} & \textbf{Study 1} & \textbf{Study 2a} & \textbf{Study 2b} \\
        \midrule
        Smartphones & 89.7 & 48.0 & 48.1 \\
        Smart Speakers & 64.1 & 42.1 & 43.4 \\
        Telephony Systems & 31.6 & {--} & {--} \\
        Desktop/Laptops  & 28.2 & 2.0 & 1.9 \\
        In-car Systems & 24.8 & 1.6 & 1.9 \\
        Touchscreen Tablets  & 22.2 & 4.3 & 3.8 \\
        Other & 1.7 & 2.0 & 1.0 \\
        \bottomrule
    \end{tabular}
\end{table}

\subsection{Study 1 Procedure}
\label{sec:procedure2}

Participants completed the survey remotely using their own device by following a link sent in recruitment materials. After following the link, participants were presented with an instruction sheet, a consent form, a demographics questionnaire and finally the main survey.
To ensure all participants coalesced on the same meaning of the term \enquote{speech interface}, the instruction sheet provided an explanation of various technologies that might constitute a speech interface:

\begin{quote}
\textbf{Speech interfaces may include a broad range of technologies, from Voice Assistants such as Amazon's Alexa, Apple's Siri, Google Assistant or Microsoft's Cortana, to various speech-based chatbots like Eviebot and BoiBot, to telephony systems used in telephone banking or ticket booking systems. 
Essentially, when referring to speech interfaces we mean \textbf{any computer system you have interacted with using speech}. 
You may have accessed these using, among other things, a smartphone, smart speaker, desktop or laptop computer, and/or in-car.} 
\end{quote}

After consenting to take part, participants then completed a demographic questionnaire asking about their age, gender and prior experience with speech interfaces,
before moving on to complete the \ac{pmq}. 
The description of what constitutes as speech interface was accessible through a mouse-over information icon throughout completion of the questionnaire. 
The survey was hosted on LimeSurvey (version 3.17.7). All 23 questionnaire items were presented in a single block, with the order in which items were presented being fully randomised between participants. 
Half the items within each factor had poles reversed in a fixed fashion between participants to ensure an even number of terms more closely related to humans (positive) and an even number most closely related to machines (negative) appeared on both the left and right side of the scale.  
After completing the \ac{pmq}, participants moved on to complete the \ac{sassi}, and then the \ac{idaq} questionnaires, which were also presented in single fully-randomised blocks. 
All questionnaire items were mandatory, meaning participants had to respond to all items in order to move on. 
Once the questionnaires were complete, participants were debriefed and thanked for their participation.

\subsection{Confirmatory Factor Analysis Results}

\subsubsection{Analysis Approach, Data Screening and Assumption Testing}
\label{sec:approach_cfa_study1}
\ac{cfa} was used to validate the factor structure identified in \cite{doyle2021we} and displayed in \autoref{tab:pca23}.
Prior to conducting analysis, data were screened for incomplete or inattentive responding using the same procedure outlined in previous work \cite{doyle2021we}.
As all items required a mandatory response, missing data was not an issue. 
However, as in \cite{doyle2021we} a number of steps were taken to ensure data quality prior to conducting the \ac{pca}. 

Common practice for assessing questionnaire data quality involves visually screening data to detect indications of \ac{ier}, such as patterned or straightline responding, with participant data omitted when evidence of \ac{ier} is detected \cite{field_discovering_2013}. However, visual screening can be error prone, particularly when screening large data sets \cite{field_discovering_2013}, and when attempting to identify dispersed patterns that use a limited set of response options (e.g., 1-2-3, 2-3-1, 3-1-2, and so on). 

With this in mind, a combinatorial screening procedure was developed with four exclusion criteria to identify patterned and inattentive responding among participants. 
Criteria, which was developed based on advice from \cite{Dunn2018irv}, included: (a) responses having an SD <0.6, (b) using the same response option for >\SI{70}{\percent} of responses, (c) using three sequential response options (i.e., 1-2-3, 3-4-5 or 5-6-7) for >\SI{90}{\percent} of responses, or (d) using a top, middle and bottom approach (i.e., responding 1-4-7) for >\SI{90}{\percent} of responses. 
Meeting criteria (a) or (b) resulted in automatic exclusion, whilst meeting criteria (c) or (d) resulted in a follow-up visual screening to aid the decision. 
In total, data from 18 participants were excluded based on this protocol, with data from 117 participants being retained for analysis. 


Tests for multivariate normality included the Mardia, Henze-Zirkler, Royston H and Doornik-Hansen tests, along with outlier detection using Mahalanobis distances. 
Although our sample size is above the minimum recommendation of 100 participants \cite{clark1995constructing, field_discovering_2013}, the Mardia, Henze-Zirkler, Royston H and Doornik-Hansen tests suggested non-normal distribution (p<.05).
Mahalanobis distances were calculated for all participants to identify multivariate outliers, with three participants’ \ac{pmq} scores being highlighted as potentially problematic (M$^2$=49.73, p<.001). 
However, removal of these outliers had only a negligible impact on multivariate normality tests and so they were retained. 
Linearity was then assessed using q-q plots, revealing a relatively linear relationship between variables. 
A cursory check showed the removal of Mahalanobis outliers also had little impact on linearity, further supporting the decision to retain this data. 

A correlation matrix was then produced, which indicated no issues with low bivariate correlations (mean r<.15) or multicollinearity (r>.9) \cite{clark1995constructing}. 
The \ac{pmq} data also produced a determinate figure of 0.00011, above the 0.00001 threshold, suggesting the data were suitable for \ac{cfa} \cite{field_discovering_2013}.
\ac{kmo} test of sampling adequacy for the \ac{pmq} overall was meritorious (0.8), with 20 out of 23 items falling in the range of meritorious (0.8-0.9) to middling (0.7-0.8). 
Exceptions with lower \ac{kmo} include item 14 (life-like/tool-like; 0.64 = mediocre); item 18 (authentic/fake; 0.68 = mediocre); and item 22 (interpretive/literal; 0.59 = miserable) \cite{kaiser1974index}. 
Results from Bartlett’s test for sphericity ($\chi^2$=973.34, p <.001) further demonstrated that the data were appropriate for confirmatory factor analysis \cite{field_discovering_2013}. Internal reliability for the measure overall ($\alpha$=.85) and factor 1 ($\alpha$=.86) was strong, moderate for factor 2 ($\alpha$=.75) but weak for factor 3 ($\alpha$=.54).  

\subsubsection{Confirmatory Factor Analysis Results}
\label{sec:cfaAnalysis2}

\begin{figure}
\caption{Path diagram of CFA output for 23-item 3-factor model, with factor loadings and between factor correlation coefficients}
\includegraphics[width=8cm]{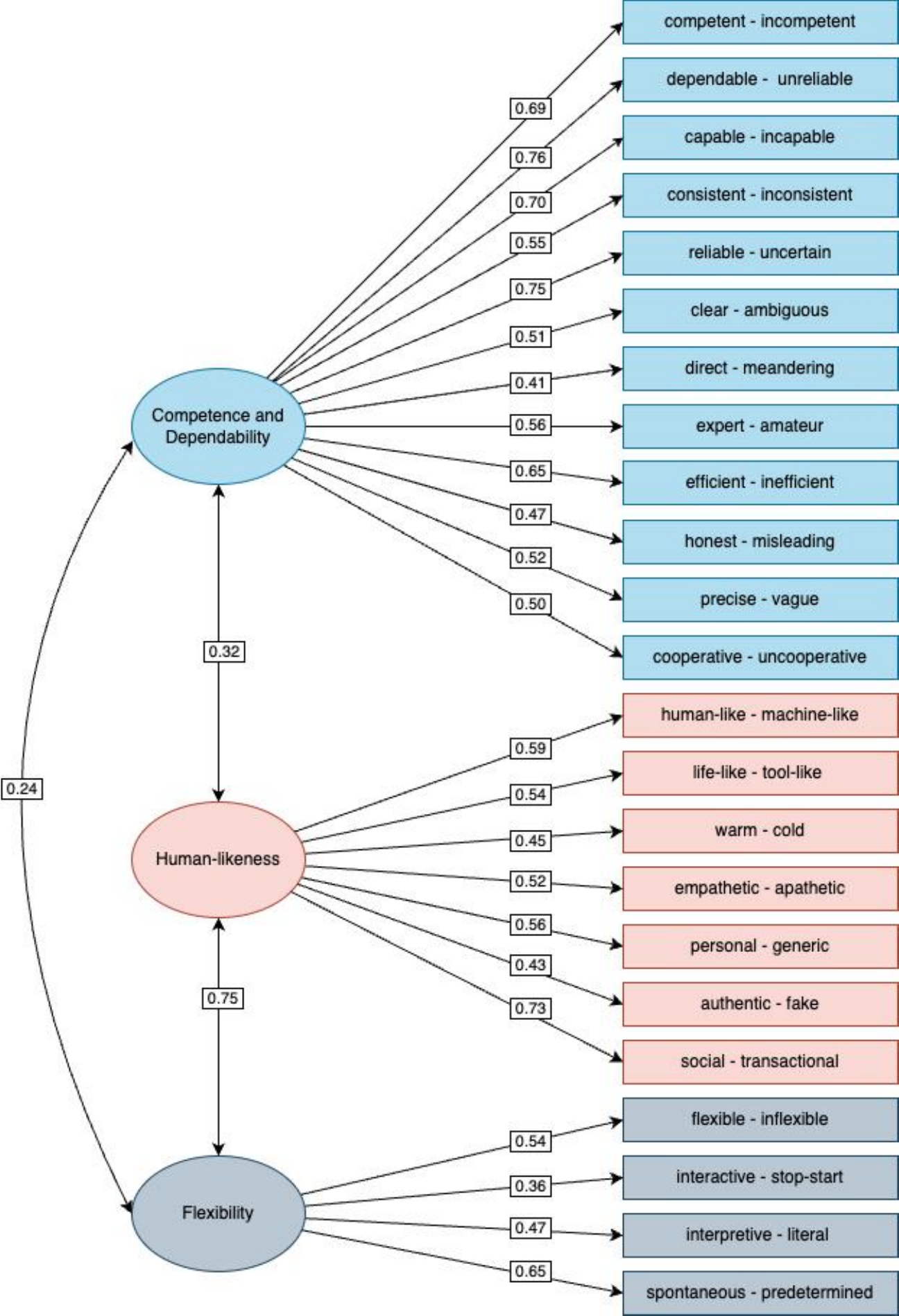}
\centering
\label{fig:23itemPMQCFA}
\Description{Path diagram of CFA output for 23-item 3-factor model, with factor loadings and between factor correlation coefficients}
\end{figure}

Following advice for dealing with non-normal data \cite{distefano2002impact,rhemtulla2012can} from smaller samples in the analysis of questionnaires with two or more factors \cite{rhemtulla2012can}, we conducted \ac{cfa} using a robust maximum likelihood estimator with Satorra-Bentler post-hoc chi-square scaling corrections \cite{satorra_scaling_1988}. Results of the \ac{cfa} are shown in \autoref{fig:23itemPMQCFA}.
In line with best reporting practices for \ac{cfa}, both robust and standard likelihood outputs are reported; however, output using robust maximum likelihood estimators are considered the most appropriate given results from univariate and multivariate normality tests. 

In accordance with best practice \cite{brown_confirmatory_2015}, statistics are also provided both for absolute and relative fit indices, alongside established thresholds for each.
Absolute fit indices include chi-square, \ac{rmsea} and \ac{srmr}. Relative fit indices include the \ac{cfi} and the \ac{tli}. 
\ac{aic} and \ac{bic} are also given, though these have no established cut-off. 
Instead, lower numbers are said to be indicative of a better fitting model \cite{field_discovering_2013}.

\autoref{tab:goodFit23} outlines \ac{cfa} output, along with established thresholds of good fit for each absolute and relative fit index. 
Although the 23-item 3-factor model appears to be relatively robust, it fell short of meeting established thresholds of good fit \cite{coughlan_structural_2008, kline_principles_2005} on three out of the five indices (\ac{srmr}, \ac{cfi} and \ac{tli}), meaning the \ac{cfa} analysis failed to fully support the hypothesised model. 


\begin{table}
\centering
\caption{Goodness of fit indices for robust and standard maximum likelihood output, with established cut-offs, for 23-item 3-factor model}
\label{tab:goodFit23}
\begin{tabular}{l S[table-format=<2.3] S[table-format=<1.2] S[table-format=<1.2] S[table-format=<1.2] S[table-format=<1.2] S[table-format=4.2] S[table-format=4.2]}
\toprule
& \multicolumn{1}{c}{Chi-square} & \multicolumn{1}{c}{RMSEA} & \multicolumn{1}{c}{SRMR} & \multicolumn{1}{c}{CFI} & \multicolumn{1}{c}{TLI} & \multicolumn{1}{c}{AIC} & \multicolumn{1}{c}{BIC} \\
\midrule
Good fit cut-offs & <.05 & < 0.08 & < 0.08 & > 0.9 & > 0.9 & {n/a} & {n/a} \\
\midrule
PMQ 23-item model (robust) & {$\chi^2$(227) = 343.20, p <.001} & 0.08 & 0.11 & 0.81 & 0.78 & 8813.36 & 8948.71 \\
PMQ 23-item model & {$\chi^2$(227) = 435.64, p <.001} & 0.09 & 0.11 & 0.74 & 0.71 & 8813.36 & 8948.71\\
\bottomrule
\end{tabular}
\end{table}

\subsubsection{Model Modification}
\label{sec:modification}

In order to further improve the structural stability of the \ac{pmq}, we examined item-level data to identify potential modifications. 
A combinatorial approach is advised here \cite{brown_confirmatory_2015}, with decisions about modifications being primarily driven by theoretical consideration of factors and items. This is then supported by inspection of factor loadings, and standardized residuals. 
Standardised residuals that indicate cross-factorial interdependence are viewed as most problematic, particularly between items and other whole factors \cite{brown_confirmatory_2015}. 

\Ac{mi} %
can also be used to support these decisions \cite{brown_confirmatory_2015}. 
However, although \acp{mi} provide the clearest evidence of problematic items, using them as the sole guide for making modifications tends to lead to the production of less reliable models \cite{brown_confirmatory_2015}; hence, theoretical guidance, factor loadings and standardized residuals should be considered first. 
With this in mind, remedial measures may be advisable in instances where interdependence between items produces an MI>10 \cite{brown_confirmatory_2015}. 
As with the analysis of residuals, modification indices that indicate cross-factorial interdependence between items, and between items and whole factors, are viewed as most problematic \cite{brown_confirmatory_2015}. 

A review of residual correlations, univariate performance, and the nature and loading of items suggested that items 6 (clear/ambiguous) and 7 (direct/meandering) from factor 1 (communicative competence \& dependability), and item 21 (interactive/stop-start) from factor 3 (communicative flexibility) were strong candidates for elimination. 
Based on the moderate negative cross-factor impact they have on the scale along with relatively weak face validity compared to other items within their respective factors, item 18 (authentic/fake) from factor 2 (human-likeness in communication), and item 10 (honest/misleading) from factor 1, were also seen as strong candidates for elimination. Similarly, \Acp{mi} also indicated that the removal of items 6, 7, 10, 18, 21, and potentially item 11 (precise/vague) from factor 1 might also improve the fit between our model and the data. 

\ac{cfa} was therefore recalculated after iteratively removing each of these items, with the aim of identifying whether removing them improved model fit. 
Items were removed in the following order based on the severity of their impact on the model's stability, according to face validity, factor loadings and residual scores: 7, 6, 21, 10 and 18. 
At this point, the model met established thresholds for good fit across four of the five indices, so item 11 was retained. 
In addition to meeting thresholds for good fit, the reduced 18-item model also exhibited strong internal reliability overall (PMQ Total: Cronbach $\alpha$=.83) and for factor 1 (communicative competence \& dependability: Cronbach $\alpha$=.85), with good internal reliability seen for factor 2 (human-likeness in communication: Cronbach $\alpha$=.74) and acceptable internal reliability seen for factor 3 (communicative flexibility: Cronbach $\alpha$=.58).

These initial findings suggested that the 18-item version may have better structural stability (see supplementary material for model fit indices), and therefore construct validity, than the 23-item version. Yet the assessment of the 18-item model involved using \ac{cfa} in an exploratory fashion, meaning the 18-item version would need to be verified through a new sample of data. The next step was therefore to conduct a further study with the purpose of directly testing the newly surfaced 18-item model and compare it to the earlier 23-item version based on a new set of responses (see \autoref{sec:study3}).

\subsection{Discussion}

Initial \ac{cfa} results suggested that the three factor structure had merit, but that a better fitting model could be achieved through minor item-level modification of the scale. 
Follow-up analysis highlighted five specific items as potential causes of instability in the model. 
Iterative removal of these items then led to the identification of a more robust 18-item three-factor model. 
Notably, the results from Study 1 can only be seen as partial confirmation of the 18-item 3-factor structure as it involved exploratory rather than confirmatory use of \ac{cfa}. We therefore ran a further study aimed at directly testing the 18-item version and comparing it to the 23-item version.

\section{Confirmatory Factor Analysis: 23- vs.\ 18-item scale}
\label{sec:study3}

Our second study was aimed at fully confirming the validity of the 18-item 3-factor version of the \ac{pmq}, and directly comparing the 23- and 18-item versions, using \ac{cfa} on data from a new sample of participants. 
For this study (Study 2), participants' PMQ ratings were gathered at three time points: an initial administration (termed Time 1); a second administration 12 weeks after Time 1 (termed Time 2); and a final administration 4 weeks after Time 2 (termed Time 3).
Follow-up administrations were conducted so we could evaluate the \ac{pmq}'s test-retest reliability.
Results for test-retest reliability can be found in \autoref{sec:reTest}. 
First, the focus is on evaluation of the \ac{pmq}'s construct validity in a comparison of the competing 23- and 18-item models using \ac{cfa}. 
Only data from the first interval (i.e., Time 1) is used in the \ac{cfa} analysis. From hereon, this is referred to as Study 2a.

\subsection{Study 2a-  Participants}
\label{sec:participants3}
In total, 254 participants completed the \ac{pmq} at this first interval (Time 1) of Study 2. 
Again, the study was given low risk ethical clearance, and was conducted according to the university's and British Psychological Society ethics guidelines. 
All participants reported being native or near-native English speakers (124 female, 127 male, 1 non-binary, 1 preferred not to say; age range: 18-72 yrs; mean age=35 yrs; SD age=12.7 yrs).

For more details about the participants in Study 2a, refer to \autoref{tab:education_combined} (educational levels), \autoref{tab:speech_interfaces_combined} (most frequently used speech interfaces), \autoref{tab:usage_frequency_combined} (usage frequency of speech interfaces), and \autoref{tab:device_types_combined} (devices used to access speech interfaces).

\subsection{Study 2a- Procedure}
\label{sec:procedure3}
The procedure followed was very similar to Study 1, with participants following a link to complete the survey remotely, using their own device. 
Participant recruitment was conducted via the Prolific platform, with the survey being built using JSPsych and hosted on a private server owned by the University. 
After following the link, participants provided informed consent and then answered a similar demographic questionnaire used in Study 1. 
In this study, we included an additional open-ended question regarding their normal speech interface usage, which participants addressed immediately before being asked to complete the \ac{pmq}. 
This was used to encourage deeper reflection on the system they used most frequently, before using the \ac{pmq} to rate that system. 
After completing the \ac{pmq}, all participants were debriefed as to the motivation of the study before receiving their honorarium, which was set at a rate of €12/hr. 
This same procedure was followed each time a participant took part (i.e., at Time 1, 2 and 3).

When administering the \ac{pmq}, all items were presented in a single block, with the order items were presented in and the reversal of item poles being fully randomised between participants.
Again, all questionnaire items were mandatory, meaning participants had to respond to all items in order to complete the study. 

\subsection{Confirmatory Factor Analysis Results}
\label{sec:cfa2}
\subsubsection{Analysis Approach, Data Screening and Assumption Testing}
Prior to analysis, data were screened for inattentive responding using the procedure outlined in Study 1 (see \ref{sec:approach_cfa_study1}). This led to the omission of data from 22 participants (N=232). 
As the aim here is to compare and validate two pre-specified models, all items must be retained, meaning screening was not carried out at an item level. 

Data did not pass multivariate normality tests, and 8 multivariate outliers were identified using Mahalanobis distances. 
Q-q plots, however, demonstrated a relatively linear relationship between data points. 
That is, while a few moderate outliers were seen at the upper tail end of the q-q plot, only negligible improvement was seen with the removal of the 8 multivariate outliers. 
These were therefore retained for analysis. 
No issues were observed in terms of low bivariate correlations (mean r <.15) or multicollinearity (r >.9).

KMO test of sampling adequacy for the \ac{pmq} overall was meritorious (0.86), with 21 out of 23 items falling in the range of middling to marvelous (0.7-0.94). 
Only human-like/machine-like (0.66) fell in the range of mediocre, though this is above the level (<0.6) where remedial action is advised \cite{kaiser1974index}. 
Results from Bartlett’s test for sphericity ($\chi^2$=1907, p <0.001), and a determinate figure of 0.000187 further demonstrated the suitability of data for conducting \ac{cfa} \cite{field_discovering_2013}.

\subsubsection{CFA Results -- 18 vs.\ 23 items}
\label{sec:cfaAnalysis}
\begin{figure}
\caption{Path diagram of CFA output for 18-item 3-factor model, with factor loadings and between-factor correlation coefficients}
\includegraphics[width=8cm]{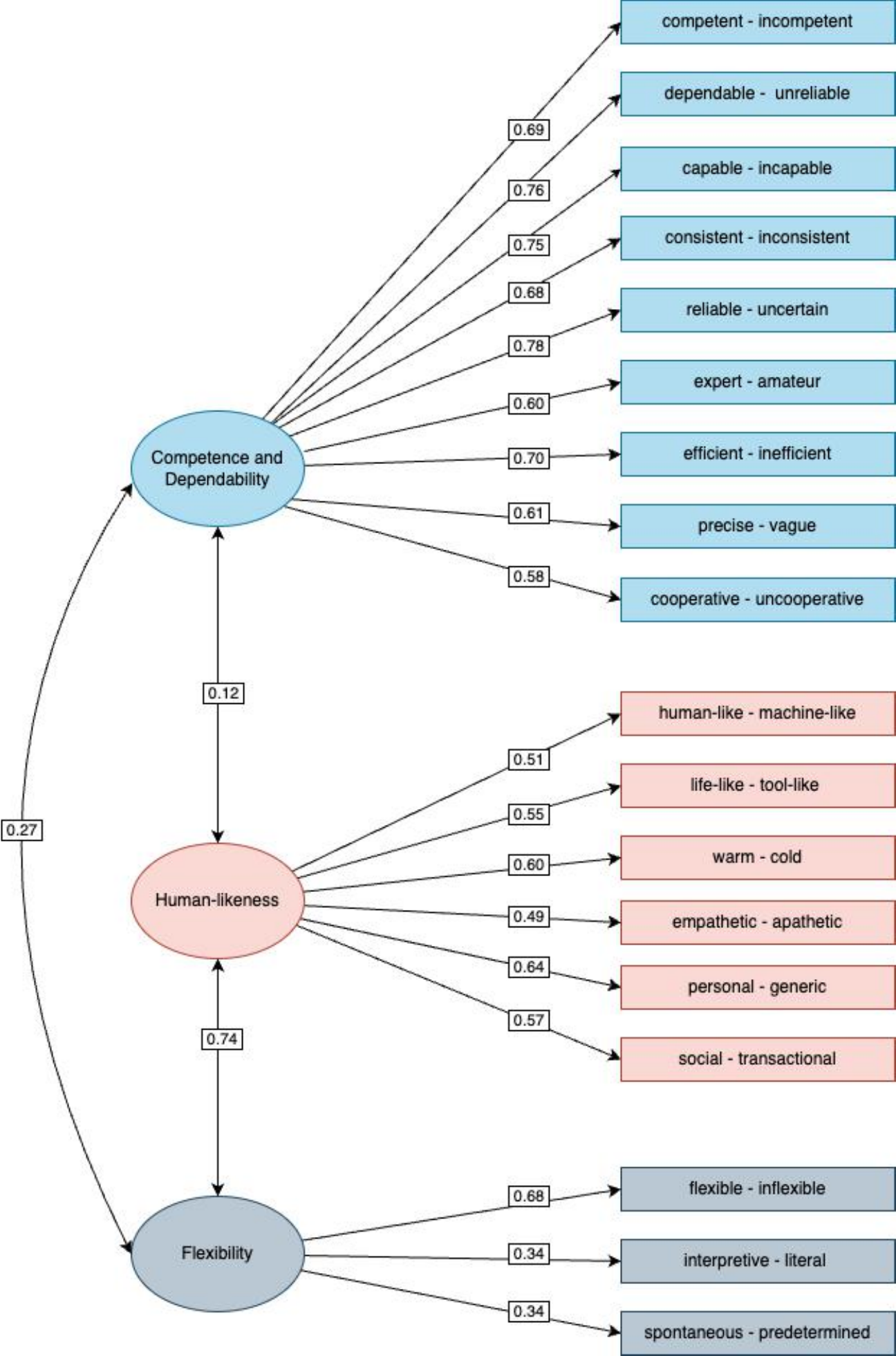}
\centering
\label{fig:18itemPMQCFAcomparison}
\Description{Path diagram of CFA output for 18-item 3-factor model, with factor loadings and between-factor correlation coefficients}
\end{figure}

\begin{figure}
\caption{Path diagram of CFA output for 23-item 3-factor model, with factor loadings and between-factor correlation coefficients}
\includegraphics[width=8cm]{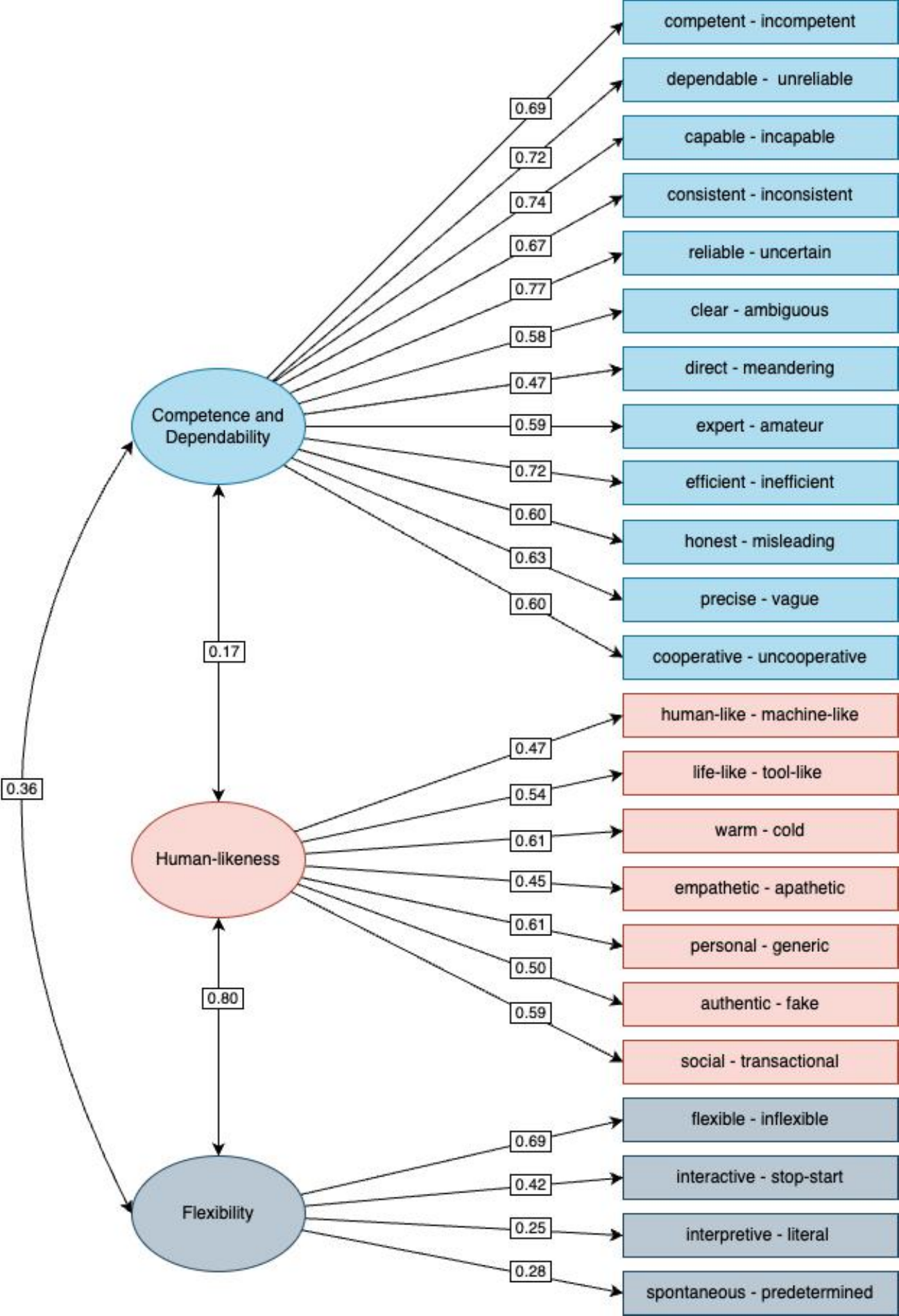}
\centering
\label{fig:23itemPMQCFAcomparison}
\Description{Path diagram of CFA output for 23-item 3-factor model, with factor loadings and between-factor correlation coefficients}
\end{figure}

Due to violation of normality tests, \ac{cfa} was again conducted using a robust maximum likelihood estimator with Satorra-Bentler post-hoc chi-square scaling corrections \cite{satorra_scaling_1988}. While both robust and standard likelihood outputs are reported (in accordance with best practice \cite{brown_confirmatory_2015}), output using robust maximum likelihood estimators indicate construct validity for the models being evaluated here. Model results for both the 18- and 23-item 3-factor \ac{cfa} are included in \autoref{fig:18itemPMQCFAcomparison} and \autoref{fig:23itemPMQCFAcomparison} respectively. 

Again, statistics are provided both for absolute and relative fit indices. 
The absolute fit indices produced include Chi-square, \ac{rmsea} and \ac{srmr}, whilst the \ac{cfi} and the \ac{tli} are our relative fit indices. 
Again, \ac{aic} and \ac{bic} statistics are also provided, with lower numbers indicating a better fitting model, rather than a hard cut-off threshold \cite{field_discovering_2013}.

\begin{table}
\centering
\caption{Goodness of fit indices for robust and standard maximum likelihood output, with established cut-offs, for 18-item and 23-item 3-factor model}
\label{tab:goodFit1823}
\begin{tabular}{l S[table-format=<2.3] S[table-format=<1.2] S[table-format=<1.2] S[table-format=<1.2] S[table-format=<1.2] S[table-format=4.2] S[table-format=4.2]}
\toprule
& \multicolumn{1}{c}{Chi-square} & \multicolumn{1}{c}{RMSEA} & \multicolumn{1}{c}{SRMR} & \multicolumn{1}{c}{CFI} & \multicolumn{1}{c}{TLI} & \multicolumn{1}{c}{AIC} & \multicolumn{1}{c}{BIC} \\
\midrule
Good fit cut-offs & <.05 & < 0.08 & < 0.08 & > 0.9 & > 0.9 & {n/a} & {n/a} \\
\midrule
PMQ 23-item model (robust) & {$\chi^2$(227) = 455.19, p <.001} & 0.07 & 0.10 & 0.83 & 0.81 & 17326.45 & 17495.34 \\
PMQ 23-item model & {$\chi^2$(227) = 564.17, p <.001} & 0.08 & 0.10 & 0.81 & 0.78 & 17326.45 & 17495.34 \\
\midrule
PMQ 18-item model (robust) & {$\chi^2$(132) = 240.76, p <.001} & 0.07 & 0.09 & 0.89 & 0.87 & 13577.93 & 13712.35 \\
PMQ 18-item model & {$\chi^2$(132) = 303.48, p <.001} & 0.08 &  0.09 & 0.87 & 0.85 & 13577.93 & 13712.35 \\
\bottomrule
\end{tabular}
\end{table}

\autoref{tab:goodFit1823} outlines \ac{cfa} output for each model, along with established thresholds of good fit across absolute and relative fit indexes. 
Internal reliability was strong for the PMQ total scale for both the 18- and 23- item versions (23-item: Cronbach $\alpha$=0.82; 18-item: Cronbach $\alpha$= .78), for both the original (Cronbach $\alpha$=0.9) and modified (Cronbach $\alpha$=0.88) versions of factor 1 (communicative competence and dependability), good for the original (Cronbach $\alpha$= 0.74) and modified (Cronbach $\alpha$=0.73) versions of factor 2 (human-likeness in communication), and acceptable for the original (Cronbach $\alpha$= 0.45) and modified (Cronbach $\alpha$=0.47) versions of factor 3 (communicative flexibility).

\subsection{Discussion}
Confirmatory factor analysis on data from Study 2 supported our initial observations in Study 1 that the 18-item, 3-factor \ac{pmq} has the better construct validity of the two models. 
Notably, the 23-item version fell below thresholds required for the scale to be described as having good construct validity, with the 18-item version showing comparatively better fit, meeting (or being closer to) thresholds for the majority of the fit indices used in analysis. Based on this, we recommend the 18-item version of the \ac{pmq} is adopted as the standard version of the scale by researchers. A full list of the 18 items and their respective factors is displayed in \autoref{tab:final18}, and can also be found alongside scoring instructions in supplementary material. Following this validation of the 18-item \ac{pmq}s construct validity, we now look to evaluate its convergent/divergent validity and test-retest reliability. 
\section{Phase 3: Convergent/Divergent Validity \& Test-Retest Reliability}
\subsection{Convergent/Divergent Validity}
\label{sec:convDiv}
\label{phase3}
Within this section, we evaluate the convergent and divergent validity of the 18-item \ac{pmq} and its underlying factors through a correlational analysis of data from Study 1. Here, correlations are used to examine hypothesised relationship between the \ac{pmq} and two other questionnaires designed to measure closely related concepts, namely: the \acf{sassi} \cite{hone2000towards}, which is designed to measure subjective perceptions of speech interface usability; and the \acf{idaq} \cite{waytz2010sees}, which is designed to measure people's general tendency to anthropomorphise non-human objects.

The \ac{sassi} consists of 34 seven-point Likert scale (1=strongly agree to 7=strongly disagree) attitudinal statements (e.g., The system is dependable) related to the subjective user experience of speech interface systems. Questions fall under six factors including \emph{system response accuracy} (9 items: 5 positively worded, 4 negatively worded; positive items reverse-scored with higher scores indicative of higher perceived system response accuracy); \emph{likeability} (9 items: all positively worded; all items reverse-scored with higher scores indicative of increased likeability); \emph{cognitive demand} (5 items: 3 positively worded, 2 negatively worded; negative items reverse-scored with higher scores indicative of higher cognitive demand); \emph{annoyance} (5 items: all negatively worded; all items reverse-scored with higher scores being indicative of higher levels of annoyance with the system); \emph{habitability} (4 items: 1 positively worded, 3 negatively worded; positive items reverse-scored with higher scores indicative of higher habitability); and \emph{speed} (2 items: 1 positively worded, 1 negatively worded; the positive item is reverse-scored with higher scores indicative that the system is perceived as being fast in the way it responds). Previous research has shown that each of the factors have strong to relatively strong internal reliability (system accuracy: Cronbach $\alpha$=.9; likeability: Cronbach $\alpha$=.91; cognitive demand: Cronbach $\alpha$=.88; annoyance: Cronbach $\alpha$=.77;  habitability: Cronbach $\alpha$=.75; speed: Cronbach $\alpha$=.69) \cite{hone2000towards}. 

The \ac{idaq} \cite{waytz2010sees} measures and predicts people’s tendency to self-engage in anthropomorphic behaviours toward non-human objects, including animals, natural entities, and technological devices. Each of these three classes has five questions related to non-anthropomorphic traits and five questions related to anthropomorphic traits (30 items in total). Advice on analysis states the 15 anthropomorphic items should be used as a single, unidimensional scale, with the 15 functional items acting as distractors to the purpose of the anthropomorphic items. This is designed to dissociate the attribution of functional and anthropomorphic traits respectively. Previous research has identified strong internal reliability of the 15 anthropomorphic items (Cronbach $\alpha$=.82) \cite{waytz2010sees}. A Cronbach alpha is not given for functional/distractor items. Each of the \ac{idaq}'s 30 items comes in the form of a question (e.g., To what extent does the ocean have consciousness?) with participants indicating their response to each statement using an 11-point scale ranging from 1 = ‘not at all’ to 11 = ‘very much’.

\subsubsection{Hypotheses}
Ideally, analysis should show \ac{pmq} factors being significantly correlated with subscales on other questionnaires designed to measure closely related concepts (convergent validity), and uncorrelated with unrelated concepts (divergent validity). Below, we outline specific hypotheses of the relationships between \ac{pmq} and \ac{sassi} subscales first, followed by those between \ac{pmq} factors and the unidimensional \ac{idaq} scale.\newline

\emph{PMQ factors \& SASSI System Response Accuracy subscale:} Conceptually, the system response accuracy subscale is closely related to notions of competence and dependability. As such, one would expect \ac{pmq} competence and dependability scores (\ac{pmq} factor 1) to be positively correlated with scores for system response accuracy (\ac{sassi} factor 1). \ac{pmq} factors 2 and 3, human-likeness and flexibility respectively, share no obvious relation with perceptions of system accuracy, though following the \ac{pmq}'s internal pattern of correlations (see \autoref{tab:pca23}) one would expect \ac{pmq} human-likeness and communicative flexibility factors to have a relationship to \ac{sassi} system response accuracy. We therefore hypothesise that all \ac{pmq} factors will have a statistically significant positive correlation with SASSI system response accuracy subscale scores (H1).\newline

\emph{PMQ factors \& SASSI Likeability subscale:} On the surface the likeability factor seems to share a somewhat mixed relationship with \ac{pmq} factors. With items that reflect how ‘pleasant’, ‘friendly’ and ‘enjoyable’ the system is, the authors describe this subscale as a measure of likeability and affect. This may suggest it is most closely related to \ac{pmq} human-likeness, which also contains some affect based items (‘warm/cold’ and ‘apathetic/empathetic’). Further, although \ac{pmq} human-likeness focuses specifically on human-likeness in terms of the way systems communicate, it is not entirely dissimilar to the concept of anthropomorphism. Research has shown that increased anthropomorphic design is associated with increased likeability of social robots \cite{salem2013err} which, given people’s general preference for anthropomorphic systems \cite{large_please_2019}, may further support the idea of a relationship between \ac{sassi} likeability and \ac{pmq} human-likeness. However, \ac{sassi} likeability also measures how ‘useful’ and easy to use a system is, how easy it is to recover from errors, and the degree of control people experience when using a system. In this sense, \ac{sassi} likeability also reflects aspects of \ac{pmq} competence and dependability, and \ac{pmq} communicative flexibility. We therefore hypothesise that all \ac{pmq} factors will share a statistically significant positive correlation with the \ac{sassi} likeability subscale (H2).\newline

\emph{PMQ factors \& SASSI Cognitive Demand:} Much like \ac{sassi} likeability, items included in the \ac{sassi} cognitive demand subscale have no direct comparison to items across the \ac{pmq}'s factors. That said, one would imagine a competent, human-like and communicatively flexible system would reduce the cognitive burden for the user. Indeed, reduction in users’ cognitive burden is one of the key reasons for designing anthropomorphic systems, under the assumption that a more natural and familiar interaction demands less thought on the part of the user regarding how to interact with a system \cite{zhang_service_2008}. As such, one would expect to see an inverse relationship between \ac{sassi} cognitive demand and \ac{pmq} factors, with higher perceived competence and dependability, human-likeness and flexibility related to lower perceived cognitive demand. We therefore hypothesise that all \ac{pmq} factors will be statistically significantly negatively correlated with \ac{sassi} cognitive demand subscale scores (H3). \newline

\emph{PMQ factors \& SASSI Annoyance subscale:} Generally speaking, one might expect a highly competent and flexible system to be less annoying to use, and anthropomorphic systems are said to be preferred over non-anthropomorphic systems \cite{large_please_2019}. Therefore, we might expect all \ac{pmq} items to share an inverse relationship with \ac{sassi} annoyance, with increased annoyance associated with less competent, flexible and human-like systems. We therefore hypothesise that all \ac{pmq} factors will have a statistically significant negative correlation with \ac{sassi} annoyance subscale scores (H4).\newline

\emph{PMQ factors \& SASSI Habitability subscale:}
\ac{sassi} habitability measures how well a user thinks they understand how to use a speech interface. Based on the author’s description of the subscale \cite{hone2000towards}, one could argue that high scores for \ac{sassi} habitability indicate a high degree of congruence between a user’s mental model for how a speech interface system works and how the system actually works. This creates an interesting comparison with the \ac{pmq}. On the surface, one might expect a positive correlation between a system that is seen as competent, human-like and more flexible, and increased perceived habitability. This is particularly true for \ac{pmq} factor 2, if one accepts the assumption that it is easier to understand how to interact with more anthropomorphic systems \cite{zhang_service_2008}. It is also true that, technically, habitability and all of the \ac{pmq} factors effectively measure a person's mental model of a speech interface system, further supporting the idea that they may be positively correlated. However, habitability was developed through the lens of functional usability, whereas \ac{pmq} factors were developed with a focus on the communicative ability of speech interfaces as social entities. \ac{sassi} habitability items are squarely focused on the degree of ambiguity users experience during interactions (e.g., “I sometimes wondered if I was using the right word”; “It is easy to lose track of where you are in an interaction with the system”), whereas the 18-item \ac{pmq} contains only two items that are broadly related to the concept of ambiguity (reliable/uncertain and precise/vague, both in factor 1). Given the nature of \ac{sassi} habitability items, and fundamental differences in terms of the traditional usability focus of \ac{sassi} and the socio-cognitive basis of the \ac{pmq}, we hypothesise that all \ac{pmq} factors will not be statistically significantly correlated with \ac{sassi} habitability subscale scores (H5).\newline


\emph{PMQ factors \& SASSI Speed subscale:}Again, there is no obvious comparison between \ac{sassi} speed and the \ac{pmq}'s factors or items. One might expect a system perceived as being slow would not be seen as being particularly competent; however, the relationship between speed and competence is not clear cut. For instance, an interlocutor may be perceived as more incompetent either because they speak too slowly or too quickly. That is, there are likely optimal and suboptimal response speeds and speech rates (i.e., just right vs. too fast/too slow) related to a dialogue partner's partner model that the  \ac{sassi} speed subscale may not adequately capture. As such, we hypothesise that all \ac{pmq} factors will not be statistically significantly correlated with the \ac{sassi} speed subscale scores (H6). \newline



\emph{PMQ factors \& IDAQ:}
The \ac{idaq} was specifically selected due to its focus on anthropomorphism, which may be related to \ac{pmq} human-likeness, but not directly to other \ac{pmq} factors. That said, given that all \ac{pmq} factors are positively correlated, it is expected that all \ac{pmq} factors will also be positively correlated with total \ac{idaq} scores. We therefore hypothesise that all \ac{pmq} factors will exhibit a statistically significant positive correlation with the \ac{idaq} (H7).\newline


\subsubsection{Participants \& Procedure}
Data used for this analysis were gathered as part of the Study 1, previously described in sections \ref{sec:participants2} and \ref{sec:procedure2}. \ac{pmq} subscale scores were calculated by reversing item poles with positively valenced adjectives relevant to each factor (e.g., competent, flexible) that were presented on the left of a semantic differential pair. Scores for each item within the subscales were then summed and averaged to create scores for each subscale that were used in the correlation analysis presented below. Full details of \ac{pmq} scoring are included in the supplementary material. \ac{idaq} and \ac{sassi} scores were calculated as per scoring instructions for those scales. Following assessment of descriptive statistics and outliers, data from 116 participants involved in Study 1 were included in this analysis. 

\subsubsection{Internal Reliability Results}
Within our sample, \ac{sassi} subscales of system accuracy (Cronbach $\alpha$=.83), likeability (Cronbach $\alpha$=.82) and cognitive demand (Cronbach $\alpha$=.74) all exhibited strong internal reliability. The remaining subscales of annoyance (Cronbach $\alpha$=.73), habitability (Cronbach $\alpha$=.61) and speed (Cronbach $\alpha$=.68) produced acceptable levels of internal reliability. Although these cronbach alpha values are lower than those reported during SASSI scale development ~\cite{hone2000towards}, this may be due to the fact that the scale was originally used to assess these concepts directly after speech system interaction, rather than asking users to assess them for a system they are familiar with, as is the case in the work presented here. The \ac{idaq}'s 15 anthropomorphic items also demonstrated strong internal reliability within the sample data (Cronbach $\alpha$=.80). As highlighted in section \ref{sec:modification}, the reduced 18-item model also exhibited strong internal reliability overall and for each factor among this sample (Communicative competence \& dependability factor: Cronbach $\alpha$=.85; human-likeness in communication: Cronbach $\alpha$=.71) with weaker internal reliability for the communicative flexibility factor (Cronbach $\alpha$=.54). 



\subsubsection{Convergent/Divergent Validity Analysis Results}
Full descriptive statistics for \ac{pmq}, \ac{sassi} and \ac{idaq} measures are included in the supplementary material. Due to violations of normal distribution within many of the subscale scores ($p<.05$), Spearman's rank correlations were calculated, assessing the relationship between each of the three \ac{pmq} factors with each of the six \ac{sassi} factors. A full list of these correlations is included in the supplementary material. A strong positive correlation was seen between \ac{pmq} competence and dependability and \ac{sassi} system response accuracy ($r_s=.76, p<.001$)\footnote{For all correlation coefficients in this section: df=114}., whilst a weak but statistically significant positive correlations was seen for human-likeness ($r_s=.26, p=.004$) and flexibility ($r_s=.23, p=.01$), supporting H1.

\ac{sassi} likeability had a statistically significant, moderate to strong positive correlation with \ac{pmq} competence and dependability ($r_s=.64, p<.001$), and significant weak positive correlations with human-likeness ($r_s=.24, p=.008$), but not for flexibility ($r_s=.14, p=.15$), partially confirming H2. 

\ac{sassi} cognitive demand shared a significant negative moderate correlation with \ac{pmq} competence and dependability ($r_s=-0.49, p<.001$) and a negative weak correlation with \ac{pmq} human-likeness ($r_s=-0.28, p=.003$), with no statistically significant relationship between cognitive demand and flexibility ($r_s=0.07, p=.47$). H3 is therefore partially supported.

\ac{sassi} annoyance was statistically negatively correlated with \ac{pmq} competence and dependability ($r_s=-0.51, p<.001$), \ac{pmq} human-likeness ($r_s=-0.42, p<.001$) and flexibility ($r_s=-0.34, p<.001$). H4 is therefore supported.

As predicted, all \ac{pmq} factors were not statistically significantly correlated with \ac{sassi}’s  habitability subscale. Therefore, H5 is supported.

\ac{sassi} speed shared a significant positive correlation with \ac{pmq} competence and dependability ($r_s=.46, p<.001$), but was not statistically significantly correlated with \ac{pmq} human-likeness ($r_s=.18, p=.06$) and flexibility ($r_s=.05, p=.53$). While partially correct, the statistical significance of the relationship between \ac{sassi} speed and \ac{pmq} competence and dependability means H6 must be rejected. 

Finally, although human-likeness exhibited a stronger relationship to \ac{idaq} total scores than the other \ac{pmq} factors as predicted, none of the relationships identified were statistically significant ($p>.05$). Therefore, H7 is rejected.


\subsubsection{Discussion}
In summary, \ac{pmq} factors shared statistically significant correlations, with \ac{sassi} subscales of system response accuracy and annoyance upholding H1 and H4. Although \ac{pmq} factors of competence and dependability, and human-likeness in communication shared a predicted negative relationship with \ac{sassi} cognitive demand and a positive relationship with likeability, the relationship between communicative flexibility cognitive demand and likeability was not statistically significant, meaning H3 and 4 were only partially supported. \ac{pmq} factors were not statistically correlated with \ac{sassi} habitability, upholding the prediction made in H5. Although \ac{pmq} human-likeness and flexibility were not correlated with \ac{sassi} speed, the competence and dependability factor was, meaning H6 is also rejected. \ac{pmq} factors were not significantly correlated to the \ac{idaq} scale either, meaning H7 is also rejected.
 
On the whole, the analysis suggests that the PMQ responses converge and diverge with scores on other similar and dissimilar scales largely as predicted. Indeed, the fact that the \ac{pmq} maps well with some \ac{sassi} factors but not others shows that it offers a distinct measure of perceptions in the context of \ac{hmd} relative to \ac{sassi}. This is likely a result of the different approaches taken in their construction. Namely, the strong usability focus of \ac{sassi} compared to the socio-cognitive concerns of the \ac{pmq}. The same can also be said of the \ac{pmq} vs. \ac{idaq} comparisons. Although a stronger association to \ac{pmq} human-likeness was expected, results suggest that the perceived human-likeness of speech interfaces is somewhat distinct from a general tendency to anthropomorphise non-human objects. This supports earlier intuitions that partner modelling of speech interface human-likeness involves appreciation for both the likeness and unlikeness of humans and speech interfaces \cite{doyle_humanness_2019, nass_does_2001}. 

\subsection{Test-Retest Reliability}
\label{sec:reTest}
The studies above confirm the structural robustness of the 18-item, 3-factor \ac{pmq} (construct validity); that it performs largely as expected against scales designed to measure similar and dissimilar constructs (convergent/divergent validity); and that it offers a distinct measure relative to those already available to researchers in \ac{hmd}. 
The following section reports on analysis aimed at quantifying the test-retest reliability of the 18-item, 3-factor \ac{pmq}. We do this through a longitudinal study wherein participants completed the \ac{pmq} at three different intervals: an initial administration (Time 1), a second administration 12 weeks later (Time 2), and a final administration 4 weeks after Time 2 (Time 3). Data used in this analysis comes from Study 2, with Time 1 data being used in the earlier \ac{cfa} to evaluate construct validity (see \autoref{sec:study3}). Essentially, the aim is to examine the extent to which measurement can be replicated \cite{koo2016guideline}, or how consistent \ac{pmq} scores are over time \cite{kline_psychometrics_2000}. Analysis will examine test-retest reliability for the whole scale and for each of the \ac{pmq}'s three factors.


\subsubsection{Study 2b Participants}

Only data from participants who completed the \ac{pmq} at all three intervals are included in this analysis. Our sample therefore included 106 participants (48 female, 56 male, 2 preferred not to say; age range 18-70 yrs, mean age= 38 yrs; SD age= 14.41 yrs).

For more details about the participants in Study 2b, refer to \autoref{tab:education_combined} (educational levels), \autoref{tab:speech_interfaces_combined} (most frequently used speech interfaces), \autoref{tab:usage_frequency_combined} (usage frequency of speech interfaces), and \autoref{tab:device_types_combined} (devices used to access speech interfaces).




\subsubsection{Study 2b - Measures and Administration Phases}
As previously mentioned, the \ac{pmq} was administered to participants at three intervals: an initial administration (Time 1), a further administration 12 weeks after initial administration (Time 2) and a final administration 4 weeks after Time 2 (Time 3). 
Using Prolific to control access, participants were only invited to subsequent administrations if they had completed the previous administration (i.e., only participants who completed the \ac{pmq} at Time 2 were invited to complete it at Time 3). 
As full validation of the 18-item version of the \ac{pmq} had yet to be completed at the time of conducting the study, participants were presented with the 23-item version at all intervals. 
However, test-retest analysis is based on the validated 18-item version only. 
As mentioned in section \ref{sec:procedure3}, when administering the \ac{pmq} to participants, the order of items, and the reversal of positive/negative poles for half of the items within each factor, were fully randomised. 
Items were then reordered prior to analysis. 
Due to data corruption, demographic variables were used to match participants across the three samples, with only exact matches in responses to age (+1), gender, education completed, occupation, nationality, IPA usage (+/-1), preferred IPA, and preferred device used to link participant responses across administrations.   

\subsubsection{Study 2b Procedure}
All details of the procedure are identical to that described in section \ref{sec:procedure3} for all administrations of the questionnaire. 

\subsubsection{Analysis approach}
Although it is common to use a correlation coefficient (Pearson's correlation), intra-correlation coefficients (\ac{icc}) are considered more accurate as they account for correlation coefficients and inter-rater agreement simultaneously \cite{koo2016guideline}. 
Both Pearson's correlation and \ac{icc} are therefore provided here, though \ac{icc} should be viewed as the more reliable indicator of re-test reliability. As no major violations of normality were observed during data screening, following advice from \cite{field_discovering_2013}, parametric Pearson's product moment correlations were used. Only participants who completed all three intervals are included in this analysis. Following assessment of descriptive statistics and outliers, the sample consisted of data from 106 participants.

\subsubsection{Internal Reliability and Test-Retest Reliability Results}
Internal reliability for the PMQ whole scale (i.e., PMQ Total- Cronbach $\alpha$ Time 1=.84; Cronbach $\alpha$ Time 2= .86; Cronbach $\alpha$ Time 3= .86), the communicative competence and dependability (Cronbach $\alpha$ Time 1= .89; Cronbach $\alpha$ Time 2= .86; Cronbach $\alpha$ Time 3= .92) and human-likeness in communication (Cronbach $\alpha$ Time 1= .79; Cronbach $\alpha$ Time 2= .84; Cronbach $\alpha$ Time 3= .85) subscales was strong, with communicative flexibility being the weakest in terms of internal reliability (Cronbach $\alpha$ Time 1= .55; Cronbach $\alpha$ Time 2= .57; Cronbach $\alpha$ Time 3= .62). Following advice from \cite{koo2016guideline}, \ac{icc} estimates and their 95\% confidence intervals were calculated using the R "Psych" package (version 2.3.3) based on a mean-rating (k=3), absolute-agreement, 2-way mixed-effects model. 
\ac{icc} coefficients were 0.86 (CI: 0.81-0.9) for PMQ total scores across the three intervals, whilst \ac{icc} coefficients were 0.79 (CI: 0.71-0.85) for communicative competence and dependability (factor 1), 0.87 (CI: 0.82-0.91) for human-likeness in communication (factor 2), and 0.80 (CI: 0.72-0.86) for communicative flexibility (factor 3). Results suggest the \ac{pmq} has good test-retest reliability as a whole scale and across each of the three factors. 


For easy comparison to other examples of test-retest reliability, Pearson's correlation coefficients are also given. 
A statistically significant positive correlation was seen for whole scale scores between Time 1 to Time 2 administrations (12-week interval) ($r =.66, p<.001$) \footnote{for all correlation coefficients in this section: df=104.}, as well as between Time 2 and Time 3 administrations (4-week interval) ($r =.77, p<.001$). 
A statistically significant positive correlation was seen for communicative competence and dependability (factor 1) scores between Time 1 to Time 2 ($r =.54, p<.001$) and between Time 2 to Time 3 ($r =.67, p<.001$). 
A statistically significant positive correlation was seen for human-likeness in communication (factor 2) scores between Time 1 to Time 2 ($r =.69, p<.001$) and Time 2 to Time 3 ($r =.80, p<.001$). 
A statistically significant positive correlation was also seen for communicative flexibility (factor 3) scores between Time 1 to Time 2 ($r =.55, p<.001$) and between Time 2 to Time 3 ($r =.59, p<.001$). Descriptive statistics for each factor across administration phases are included in supplementary material. The findings thus show strong test-retest reliability for the 18-item PMQ when used as a whole scale (PMQ Total) and for each of the subscales.


\subsubsection{Discussion}
Our findings suggest that the 18-item \ac{pmq} scale shows strong test-retest reliability at 12-week and 4-week gaps between administration. 
This is shown both for unidimensional use of the \ac{pmq} (e.g., \ac{pmq} total scores) and for each of its subscales. As with all test-retest reliability evaluations, our efforts were reliant on participants volunteering to repeat the study, hence self-selection bias should be taken into account when interpreting findings. Nonetheless, given the broad psychological nature of the scale \cite{watson2013effect} and the use of semantic differentials \cite{kline_psychometrics_2000}, these findings are a strong indicator of robust test-retest reliability of the 18-item PMQ.
Further, researchers can expect even better test-retest reliability at shorter intervals between administrations \cite{kline_psychometrics_2000}. Results here should give \ac{hmd} researchers confidence in using the \ac{pmq} in longitudinal work, particularly given most longitudinal work in \ac{hmd} research tends to cover periods ranging from two weeks \cite{lovato_hey_2019, oh_differences_2020} to a month \cite{beneteau2019communication, beneteau_assumptions_2020, pfeifer_is_2011, porcheron2018voice}.
While few studies extend beyond this \cite{bentley_understanding_2018, cho_once_2019}, the availability of a scale with demonstrated test-retest reliability over a twelve-week interval might also encourage \ac{hmd} researchers to engage in more extended longitudinal work.

\section{General Discussion}
Although research within speech interfaces from a user perspective has been growing, there is a need for well-validated self report measures to assess concepts related to speech interface perception \cite{clark2019state}. 
Recent work on understanding language production processes in human machine dialogue has coalesced around the concept of partner models \cite{doyle2021we}, with the aim of understanding and identifying the impact of people's perceptions of machines as communicative partners on interaction \cite{branigan_role_2011, cowan_whats_2019, moore_appropriate_2017}. 
Our paper contributes by developing and validating an 18-item multi-dimensional self report scale as a way to measure and observe the concept of partner models in \ac{hmd} research.  
Across the three studies conducted, our work shows that the self-report scale developed (termed the Partner Modelling Questionnaire, or \ac{pmq}) is constructed of three factors. 
The first of these, \textit{perceived competence and dependability in communication}, stems from a focus on communicative attributes including, but not limited to: competence, dependability, reliability, consistency and efficiency. 
The next dimension is \textit{human-likeness in communication}, which stems from broad human-machine comparisons, perceptions of a system’s capacity for warmth and empathy, and how social-transactional interactions feel. 
Finally, the work identified \textit{perceived communicative flexibility} as an important partner model dimension that stems from a concern with how flexible or spontaneous a system appears to be in dialogue, and its capacity for interpretation. 
Through the studies presented, we also evidence the validity and reliability of the scale, showing that the \ac{pmq} and its subscales are robust in terms of factor structure (Section \ref{sec:study3}), internal reliability (Sections \ref{sec:modification}, \ref{sec:cfaAnalysis}), test-retest reliability (Section \ref{sec:reTest}), and convergent/divergent validity (Section \ref{sec:convDiv}).
The \ac{pmq} also provides a metric that can be used to assess the partner model concept more directly than is currently possible, meaning researchers no longer have to rely exclusively on language production effects and highly constrained research paradigms in making assertions about partner models for speech interfaces.


\subsection{The multidimensionality of partner model measurement} 
Our work offers a multidimensional tool to measure partner models, allowing researchers to measure user perceptions on three key factors. 
The factors identified within the \ac{pmq} are broadly in keeping with how the concept is described in previous work, with communicative ability and social relevance of an interlocutor being seen as critical to the concept \cite{branigan_role_2011, brennan_two_2010, clark_using_1996}. 
Findings related to the importance of perceived competence and dependability are also in keeping with the focus of partner modelling research in terms of the assumed role they play in facilitating communicative success. 
Likewise, the importance of human-likeness in communication maps to work that suggests design choices such as accent \cite{cowan_whats_2019} and anthropomorphic dialogue strategies \cite{brennan_effects_1994} significantly influence partner modelling of systems and that this may have a subsequent impact on user behaviour \cite{branigan_role_2011, cowan_whats_2019}. 
The multidimensional model produced here suggests partner models in \ac{hmd} are more detailed and complex than more general explanations offered in prior work reflect. 
Analysis carried out here also builds on earlier work where the multidimensional nature of the partner models was initially identified \cite{doyle2021we}, by demonstrating the validity and robustness of the dimensions discovered, and in quantifying the scale's reliability over time. 
This is not to say there are no other important dimensions of partner models to consider in the context of \ac{hmd}, nor that the \ac{pmq} is a monolith for measuring partner models.
However, what the \ac{pmq} does provide is the most parsimonious account of partner models for speech interface dialogue partners, and the most valid and reliable tool for measuring them at present.
Crucially, this now allows researchers to quantify the potentially nuanced effect of system design and behaviour on these specific dimensions of user perception, rather than assuming specific design elements are universally impactful, or only impact user perceptions as intended (e.g., that design changes meant to enhance human-likeness will have no unintended impact on perceived competence). 


\subsection{Interpreting scale reliability and validity} 
Although reliability and validity analysis suggests there are areas where the scale might be improved, findings from the factor analysis and repeated administration across several studies indicate that the 18-item, three-factor \ac{pmq} has relatively strong internal reliability and a robust underlying factor structure. 
We also demonstrate good test-retest reliability both at 12- and 4-week intervals between administrations. 
Results also suggest the \ac{pmq} overlaps and diverges from other similar scales available to \ac{hmd} researchers, meaning it performs largely as expected and offers a conceptually unique measure of user perceptions in this context. 


Communicative flexibility (factor 3) was the weakest of the factors in terms of internal reliability, and its removal from analysis had little to no impact on \ac{cfa} output.
As such, it could be argued that this factor is somewhat superfluous, and could potentially be eliminated.
However, communicative flexibility is a key dimension of dialogue capability, it was a key concern among participants in our original study, and interest around this issue is growing across \ac{hmd} research with the advent of LLM-supported conversational agents and the additional communicative flexibility they exhibit. It is also true that the poor internal reliability it demonstrates is likely an artefact of the small number of items (three) contained within it, rather than anything inherently unreliable about the concept and its items, hence our decision to retain it and attempt to reinforce this factor with additional items in future work. 
This decision is further supported by the strength of factor 3's test-retest reliability, across both 4- and 12-week intervals.


The \ac{pmq} also performed largely as expected in comparisons with similar and dissimilar factors on established measures. Strong correlations existed between \ac{pmq} competence and dependability and \ac{sassi} system response accuracy, cognitive demand, annoyance, likeability and speed susbscales, suggesting perceived competence and dependability is closely associated to these concepts. The lack of a relationship between \ac{sassi} habitability and \ac{pmq} factors was also as predicted. Despite both concepts being broadly associated with the concept of mental models, they do approach the concept from different angles and use different forms of measurement. While \ac{sassi}'s habitability scale focuses on potential \emph{interaction} outcomes from the formation of a user's partner model (e.g., \enquote{I sometimes wondered if I was using the right word.}), the PMQ focuses more squarely on perceptions of the system as a dialogue partner, capturing the user's description of the \emph{system}. Indeed, the difference in these concepts is also reflected in how they are measured. Habitability seems more appropriately measured through attitudinal statements whereas user reflections of their perceptions of the capabilities of a machine dialogue partner seem more suited to semantic differential scales that allow comparison and relative assessment on specific dimensions. That said, while the negative correlation between \ac{pmq} factors and \ac{sassi} cognitive demand was as expected, the lack of statistical significance for communicative flexibility was somewhat surprising, as one would imagine that speech interfaces that are perceived to be less flexible communicators would also elicit perceptions of higher cognitive demands from an interlocutor. Being forced to use a highly constrained set of commands to produce a variety of output might be perceived to be quite demanding, or at least frustrating. Alternatively, it could be hypothesised that interacting with a less flexible speech agent may be perceived as less cognitively demanding because fewer response options are permissible (i.e., the user has fewer commands to think about). Future work might look into each of these possibilities, potentially using both physiological and subjective measures to determine optimal and sub-optimal levels of system flexibility in terms of impact on a user's cognitive burden. More generally, these findings, along with the key role cognitive demand plays in explanations about when people are likely or unlikely to utilise partner models in dialogue, would suggest the relationship between partner modelling and cognitive demand warrants further examination in future work.  

The remaining \ac{sassi} factors and the \ac{idaq} were largely unrelated to \ac{pmq} factors, indicating substantial conceptual divergence between the scales. 
This was particularly interesting in relation to the \ac{idaq} as it suggests \ac{pmq} human-likeness (or perceived human-likeness in the context of \ac{hmd})  is at least qualitatively distinct from our general tendency to anthropomorphise non-human objects. 
Whether this is because \ac{pmq} human-likeness measures how human-like/machine-like a system is perceived to be, or whether this is due to the \ac{pmq}s specific focus on communicative ability, remains to be seen. 
One exception in terms of conceptual divergence between the measures is the statistically significant relationship between \ac{pmq} competence and dependability and \ac{sassi} speed, suggesting a relationship between speed of interaction response and competence and dependability judgments. Such a relationship makes sense in that speech interfaces that are faster are also seen as more competent and dependable communication partners. However, as we state when outlining our hypotheses in Section 7.1.6, this relationship may not be straightforward and warrants further work to identify the causal links between speed and competence judgements.

\subsection{Research Considerations and Future Directions}
As highlighted above, our work provides a much needed self-report scale that can now be used to more directly measure partner models in the context of speech interface interaction. 
The fact that the \ac{pmq} offers a reliable and validated tool that is cheap and easy to administer opens multiple avenues for \ac{hmd} work aimed at better understanding the association between system design, user perceptions, user behaviour and improving the user experience. 
Here we take the opportunity to highlight key considerations researchers need to make when applying and interpreting findings from the \ac{pmq}, and potential avenues for future work. 
Specifically, we elaborate on (i) the need to understand how many dimensions might be active when partner models are elicited; (ii) partner model sensitivity to design, (iii) how adaptive partner models are over time, (iv) how the \ac{pmq} relates to language based measures and effects, and (v) how widely applicable the \ac{pmq} is to other types of conversational user interfaces and conversational partners. We note that these are by no means exhaustive, but outline a set of potential avenues for future work.

\subsubsection{Activation of partner model dimensions in interaction}
Similar to the idea that the \ac{pmq} should not be considered a monolith, it also should not be regarded as a rigid, or fixed representation of partner models. 
Indeed, it is unlikely that all the dimensions presented in the \ac{pmq}s 3-factor model are activated at all times during interaction. 
The exact number of dimensions that might be simultaneously active during an interaction is currently unclear, but previous work on mental models suggests these are likely to be limited to representation of key salient features \cite{johnson-laird_mental_2010, norman1983some}. 
We propose that certain design features and system behaviours will both increase and decrease salience of some dimensions over others. 
For instance, certain dimensions may be more or less salient in different use cases (healthcare vs. e-commerce, or public vs. private use), and among different cohorts who may have different experiences on which they build their partner models. 
Indeed, the \ac{pmq} offers researchers a validated tool for quantifying key differences of this nature that is both cheap and easy to administer. 

\subsubsection{Sensitivity to design changes}
The \ac{pmq} also now enables researchers to gather deeper and more nuanced insights into how specific aspects of system design (e.g., robotic vs.\ human-like voices \cite{moore_appropriate_2017}; accent \cite{cowan_whats_2019}) and experiences in interaction (e.g., errors or miscommunication) may influence partner models and their underlying dimensions. 
Furthermore, the \ac{pmq} could also be used to quantify the effect that modifications to speech synthesis or dialogue management architectures might have on partner models, and in observing whether effects are influenced by interaction domain, situation, or are cohort dependent. 
Importantly, the \ac{pmq} allows researchers to test and quantify these potential issues and intuitions at a higher level of detail and acuity than was previously possible, whilst also offering continuity in how the concept is defined and measured across studies. 

\subsubsection{Adaptation of partner models over time}
Further conceptual work also needs to be conducted on how partner models are shaped over time. 
Some accounts of partner modelling in \ac{hhd} propose that early \emph{global} partner models are initially derived from minimal cues early in interaction. However, when beneficial to do so, these global models are updated based on interaction experiences as people work toward developing a more accurate, individualised \emph{local} partner model for a specific dialogue partner \cite{brennan_two_2010}. 
Yet, it is currently unclear how we form our initial global models for machines as dialogue partners, or what types of experiences might encourage us to re-evaluate and develop a local partner model of a specific system.
Work needs to specifically prioritise investigating how partner models are formed and adapted in accordance with interaction experiences and partner design.
It might also look to investigate what other factors may influence the transition from global to local models, and how these models may change over time. 

\subsubsection{Relationship between partner models and language production in HMD}
The development of the \ac{pmq} also allows researchers in \ac{hmd} to decouple the measurement of partner model effects from whether users react linguistically to specific events or partner characteristics in dialogue. 
As highlighted in section  \ref{sec:partnerModelMeasurement}, partner models are currently not measured directly, with alignment \cite{branigan_role_2011, cowan_does_2015} or audience design \cite{cowan_whats_2019, oviatt_linguistic_1998} effects used as proxy measures to allude to an influence and/or change in partner models in \ac{hmd}. 
The ability to now decouple partner model measurement from language effects is a significant contribution of this work, meaning researchers can now better gauge the role partner models play in user interaction behaviours in this context. 
We suggest that future work should use the \ac{pmq} to untangle and explore the influence partner models have on language production and audience design processes in \ac{hmd} interactions.

\subsubsection{Relevance to other Conversational User Interfaces and other conversational partners}
Given the way the \ac{pmq} was developed, it can reliably be applied in research investigating perceptions of any disembodied speech interface systems that communicate using voice. 
This includes general purpose systems such as \ac{ipa}s and specialised systems designed for carrying out specific tasks (e.g., an automated order taker, or automated customer service agent). 
Indeed, the \ac{pmq} is well placed to investigate the key differences in how general-purpose systems, systems designed for social interaction, and specialised task-oriented systems are perceived. 
Based on this, we encourage future work to focus specifically on how the \ac{pmq} might also be applicable to other types of dialogue systems, such as text-based disembodied agents (e.g., chatbots like ChatGPT), and embodied conversational agents such as social robots or systems that use avatars to represent the agent. 
It may be the case that partner models for speech-based agents compared to text-based agents, or for disembodied agents compared to embodied agents (i.e., avatar-based systems and social robots) are not fundamentally different to those identified in our work, but that certain \ac{pmq} dimensions are prioritised over others in the respective scenarios. 
For instance, according to previous work, the use of human language is enough to provoke comparisons between humans and machines \cite{nass_computers_1994, nass_machines_2000} but the tendency appears to be heightened in speech-based interactions \cite{gong_when_2007}. 
The importance or salience of perceived human-likeness may therefore be somewhat diluted in text-based interaction relative to speech-based interaction. 
Further, perceived animacy is said to be integral to interactions with embodied conversational agents~\cite{bartneck_measurement_2009, laban2021}, but may not be influential in interactions with disembodied systems due to their lack of physical form, which also results in substantial relative deficits in their capacity for non-verbal communication (e.g., backchanneling through hand gestures and facial expressions). 
Yet, recent work has speculated that text unfolding on screen may also elicit perceptions of animacy ~\cite {laban2021}, suggesting this may also be an important perceptual dimension in text-based interactions. 
This may highlight a need to add further items to cover concepts within partner models that are present when interacting with text and embodied speech agents, or even a need to develop new versions of the \ac{pmq} suited specifically for these different types of automated dialogue partner. What is more, for similar reasons to those outlined above, we would suggest the \ac{pmq} is not currently appropriate to administer as a measure of perceptions toward human dialogue partners. Given comparisons to humans are a common feature of \ac{hmd} research, it would be beneficial for future work to either test the current metric with human partners or use the \ac{pmq} as a base for developing a reliable measure for this purpose, rigorously testing its relationship to the \ac{pmq}.


\section{Limitations}

Our findings show the \ac{pmq} has strong validity and reliability, yet there are still areas where reliability and validity may be improved, and additional validity concepts that remain to be assessed.
As previously mentioned, factor 3 (communicative flexibility) was consistently the worst performing aspect of the scale, which may be improved with the inclusion of additional items.
Future work should aim to identify additional items that might improve the reliability and validity of this dimension and, by extension, the scale overall. 
Extending rather than eliminating this factor also carries a two-fold benefit of offering more detailed information and improving the imbalance in the number of items contained within each factor, which can sometimes cause instability within a model \cite{kline_handbook_2013}. 
That said, due to the interdependence between items, the addition or elimination of further items would need to be carefully considered and rigorously tested before being used in applied research. What is more, when considering scale reliability, future work should look to identify how the context of administering the PMQ scale (e.g. to assess a direct interaction compared to an imagined or previous interaction) may influence subscale internal reliability.

Throughout our analysis, sample size was repeatedly checked for adequacy, with samples falling within the bounds recommended for conducting \ac{pca} and \ac{cfa}. Despite this, a larger sample size needs to be considered in future work as this may produce clearer models and normally distributed data. We note that achieving normally distributed questionnaire data is known to be a challenge due to the nature of questionnaire data \cite{brown_confirmatory_2015} and the broadness of the psychological concepts being measured \cite{kline_psychometrics_2000}) 
However, it is still strongly recommended that future studies aimed at further assessing reliability and validity or further developing the \ac{pmq} aim to recruit large samples for such work, comparing our findings to those found with such samples.

Our work attempted to evaluate the \ac{pmq}'s validity, focusing on construct validity (through \ac{cfa}), and convergent/divergent validity (through assessing its relationship with two other relevant questionnaires). However, the assessment of convergent/divergent validity was limited by the small number of contextually relevant, established measures available for comparison. We selected \ac{sassi} and \ac{idaq} because they were designed to measure perceptions of non-embodied agents, they appeared to map conceptually to the \ac{pmq} factors, and their underlying structure had been demonstrated using factor analysis. Although designed specifically for \ac{hri}, the Godspeed questionnaire \cite{weiss2015meta} has consistently demonstrated strong internal reliability, including across eleven translations \cite{Bartneck2023} and recent work has shown that it has a strong underlying structure \cite{Szczepanowski/etal:2023}, making it an appealing option for comparison in future work. 
Future work should also look to evaluate the \ac{pmq} across other forms of validity. Crucially, research should be carried out that highlights the scale's discriminant validity to ensure it is sensitive enough to capture potential differences in designs and experimental conditions. Recent findings \cite{doyle2022dimensions} suggest that the measure is effective in this regard, showing sensitivity to differences between groups when a system commits more or less errors. Other validity concepts such as the \ac{pmq}'s ability to consistently predict behaviour and/or scores on other tests or scales (i.e., predictive validity) should also be explored in future work aimed at building on the validity findings in our current study.

\section{Conclusion}
Users' partner models are thought to significantly influence user interaction, yet prior to work presented here, no metric existed to directly measure the concept. 
Through the three studies presented, we have described the stages of standardisation and validation of an 18-item, 3-factor self-report semantic differential scale designed to measure people's partner models of speech interfaces. 
The work contributes a robust, reliable and valid three-factor scale for measuring the concept of partner models in the context of \ac{hmd}, refereed to as the Partner Modelling Questionnaire, or \ac{pmq}. 
As highlighted in our discussion, we hope that this measure can be a catalyst for much-needed research on how system design and interaction influence people's perceptions of speech interfaces as dialogue partners, as well as key foundational research on the concept and dynamics of partner models more generally.

\bibliographystyle{ACM-Reference-Format}
\bibliography{2022-PMQ-article/betterBib} 

\section{Author statement}

The current manuscript is not under concurrent consideration and has not been previously submitted to a conference or journal. Content within the submission has been previously published. Content describing the item pool generation, scale construction and principal component analysis (Section 3) are summaries of work conducted in \cite{doyle_humanness_2019} and \cite{doyle2021we}, respectively. The current manuscript builds on this research, fully constructing and validating a questionnaire based on this previous work. Here we execute and report on a further two novel unpublished large scale studies to a) assess PMQ construct validity through confirmatory factor analysis (here Section 4, Study 1), b) directly comparing the construct validity of 18-item and original 23-item factor structures (here Section 5, Study 2), whilst also assessing c) convergent/divergent validity (here Section 6, Study 2) and d) test-retest validity across 12- and 4-week administrations of the 18-item measure (here Section 7, Study 3).

\end{document}